\documentclass[prx,twocolumn,superscriptaddress]{revtex4-1}
\usepackage{graphicx}
\usepackage{amsmath}

\begin{document}
\title{Crystal growth in confinement} 
\author{Felix Kohler}
\affiliation{The NJORD Centre, Dept of Physics, University of Oslo, P.O.box 1048 Blindern, 0316 Oslo, Norway}
\author{Olivier Pierre-Louis} 
\affiliation{Institut Lumi\`{e}re Mati\`{e}re, UMR5306 Universit\'{e} Lyon 1 - CNRS, 69622 Villeurbanne, France}
\author{Dag Kristian Dysthe}
\affiliation{The NJORD Centre, Dept of Physics, University of Oslo, P.O.box 1048 Blindern, 0316 Oslo, Norway}

\begin{abstract}
The growth of crystals confined in porous or cellular materials is ubiquitous in Nature and industry.  Confinement affects the formation of biominerals in living organisms, of minerals in the Earth's crust and of salt crystals damaging porous limestone monuments, and is also
used to control the growth of artificial crystals.  However, the mechanisms by which confinement alters crystal shapes and growth rates are still not elucidated.  Based on novel \textit{in situ} optical observations of (001) surfaces of NaClO$_3$ and CaCO$_3$ crystals at nanometric distances from a glass substrate, we demonstrate that new molecular layers can nucleate homogeneously and propagate without interruption even when in contact with other solids, raising the macroscopic crystal above them.  Confined growth is governed by the peculiar dynamics of these molecular layers controlled by the two-dimensional transport of mass through the liquid film from the edges to the center of the contact, with distinctive features such as skewed dislocation spirals, kinetic localization of nucleation in the vicinity of the contact edge, and directed instabilities. Confined growth morphologies can be predicted from the values of three main dimensionless parameters.
\end{abstract}

\maketitle

Living organisms grow crystals to form bones, shells, coccoliths and other complex biominerals. 
Confinement of these biogenic crystals during their formation allows for fine-tuned control of ionic composition, pH and supersaturation~\cite{Mann2001} and permits to restrict growth and to define crystal shapes~\cite{Park2002,Aizenberg2003}.
Recent strategies for biomimetic materials design focus on confinement of growing crystals by complex templates to control crystal morphology and strength~\cite{Meldrum2008,Aizenberg2010}. 
In the Earth's crust minerals also crystallize in confinement during diagenesis and metamorphism~\cite{Gratier2013}. 
Furthermore, confinement is used to control crystallization of ice~\cite{Algara-Siller2015},  proteins~\cite{Huber2015}, micro- and mesoporous crystals~\cite{Fan2008} and low dimensional nanostructures~\cite{Gao2019}. Nevertheless, the  understanding of the microscopic mechanisms by which nanoconfinement controls
the morphology is still lacking.

As imaging methods reach higher resolution, our view of crystal growth is evolving~\cite{Li2012,Nielsen2014,Algara-Siller2015,Yoreo2015,Ma2020}.  However, nanoconfined crystal growth remains largely unexplored because high resolution measurement techniques such as scanning tunneling microscopy, atomic force microscopy~\cite{Ma2020} and traditional electron microscopy can only image open surfaces. Recent development of liquid cell electron microscopy has allowed the imaging of nucleation and growth of nanoparticles~\cite{Li2012,Nielsen2014,Yoreo2015} and confined ice~\cite{Algara-Siller2015}, but cannot image atomic step dynamics in confinement. Optical microscopy has been used with atomic scale resolution for studies of 
unconfined crystal growth from solution {\em in situ}~\cite{Bunn1949a,Tsukamoto1983,Sazaki2010}. In  nanoconfinement, reflection interference contrast microscopy (RICM) has previously reached a measurement precision of 2-30~nm~\cite{Theodoly2010}.

We have used RICM with high intensity LED illumination, high  resolution camera and image analysis to achieve for the first time sub-nanometer resolution topography measurements of atomically flat NaClO$_3$ crystals growing in nanoconfinement. Our measurements allow us to analyze quantitatively the growth process and unravel a nanoconfinement regime that drives standard features of crystal growth into {\em behaviors that are distinct from those of free surfaces}~\cite{Saito1996,Misbah2010}. Among these peculiar behaviors of nanoconfined growth, we observed skewed spirals, strong localization of nucleation along the contact edge, and instabilities such as fingering and bunching of molecular steps dictated  by the orientation of and distance to the contact edge. Confined growth therefore proceeds with a specific growth mode characterized by two-dimensional transport of growth units along the liquid film from the edge to the center of the contact, which produces gradients that control the growth of new molecular layers.
\begin{figure*}
    \centering
    \includegraphics[width=16cm]{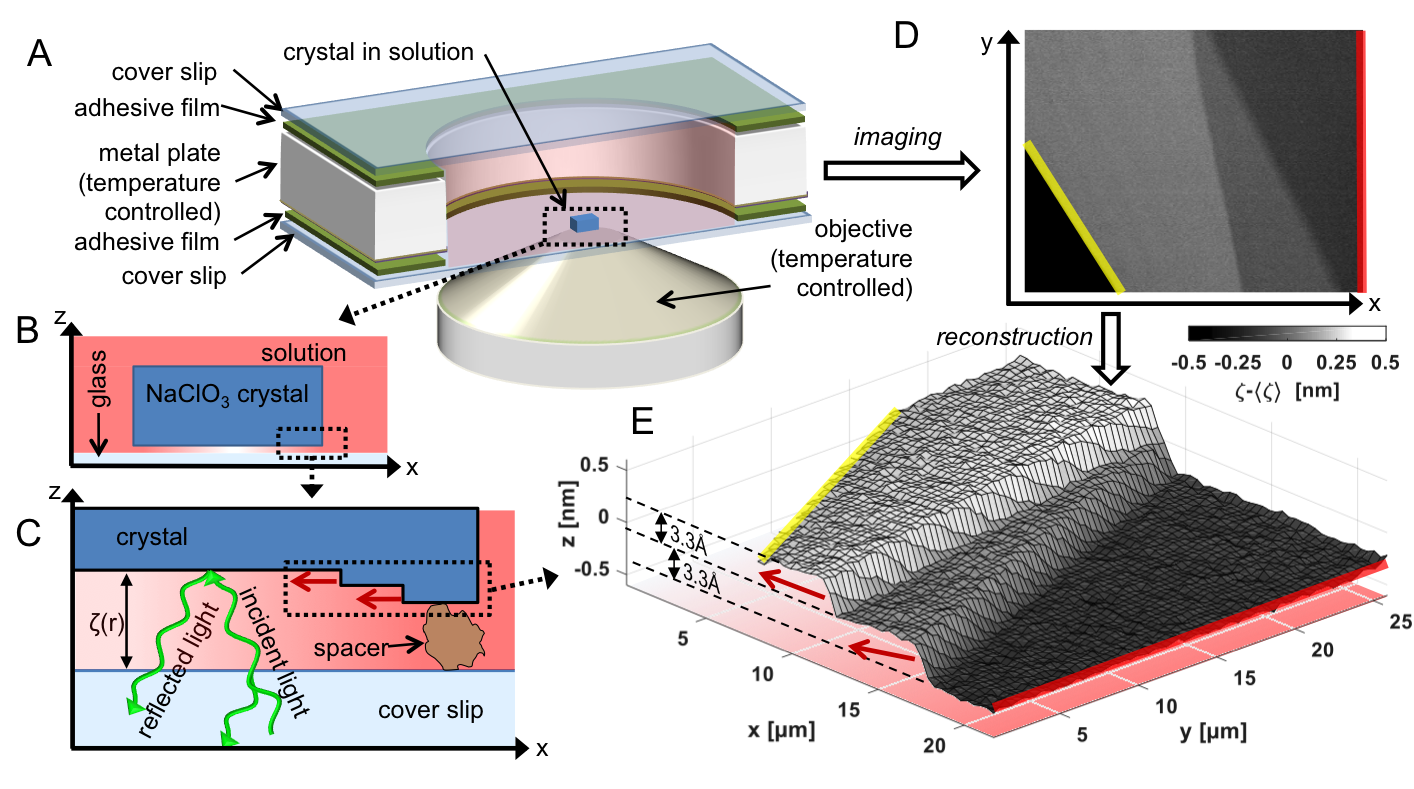}
    \caption{{\bf Experimental setup}
{\bf{A}} Sketch of experiment chamber with crystal in solution and high resolution microscope objective. 
{\bf{B}} Vertical cut of crystal. 
{\bf{C}} Reflection of light from crystal and confining surface enables interferometric determination of 
the distance $\zeta(r)$ between the crystal and the glass surface and schematic of molecular steps of layered growth from outer edge.
{\bf{D}} Interferometric image of part of a growth rim of a crystal surface. 
{\bf{E}} Reconstruction of local crystal height relative to the time averaged interface $z=\zeta-\bar{\zeta}$, with $\bar{\zeta}=53$~nm from the image intensity in {\bf{D}}. The measured 0.33~nm height of the steps corresponds to single molecular layers of the NaClO$_3$ crystal. In subfigures {\bf{B, C}} and {\bf{E}} the colour intensity indicates the solution supersaturation from high (red) in the bulk to zero (white) at the center. The red arrows indicate the growth direction of the steps as interpreted from the time lapse movies S1-S5~\cite{MatMet}.}
    \label{fig:setup}
\end{figure*}

\section*{Methods}
We observe single NaClO$_3$ and CaCO$_3$ crystals growing from solution with supersaturation $\sigma$ precisely controlled. The CaCO$_3$ crystal growth has been described in detail earlier~\cite{Li2017b,Li2018_CGD} and we will here describe the NaClO$_3$ experiments.
As shown in Fig.1~A, the nanoconfined crystal growth is observed from the bottom using RICM. The glass cover slip shown on Fig.~1~A-C supports the weight of the crystal 
and acts both as confining interface and as the reference mirror. The light reflected from the crystal and from the cover slip interfere and the intensity of the resulting image is a periodic function of the distance $\zeta$ between the two surfaces.  For distances $\zeta<125$~nm the height
can be calculated from the image intensity with unprecedented, sub-nanometer precision~\cite{MatMet} shown in Fig.~1~D\& E. This height profile reconstructed from the RICM image shows two step fronts at the edge of growing atomic layers on the (001) facet
with a thickness of 0.33~nm each.
\subsection*{Sample Preparation}
The sample is prepared under a laminar flow bench using pre-cleaned cover components. The chamber is filled with a defined (50~$\mu$l) volume of saturated solution. A seed crystal of approx. 0.5~mm$^3$ is added to the solution before the chamber is sealed. Before the actual experiment, the crystal is dissolved to a diameter of approximately 50~$\mu$m. This will dissolve all other small crystals in the chamber, which may accidentally appear and which can act as a seed of further crystallization processes. The high nucleation barrier of NaClO$_3$ impedes the spontaneous nucleation of additional crystals during the experiment.

Because of the solubility of NaClO$_3$ is $10^4$ times higher than CaCO$_3$, the Debye screening length and crystal-glass distance $\bar{\zeta}$ at equal pressure is much smaller for NaClO$_3$ than for CaCO$_3$. In order to obtain a similar crystal-glass distance $\bar{\zeta}$ in the two sets of experiments a layer of spacer particles of diameter 10-80~nm is dispersed between the glass coverslip and the NaClO$_3$ crystal~\cite{Kohler2018} (see Fig.~S2~\cite{MatMet}).

\subsection*{Temperature and imaging control}
The saturation concentration, or solubility, $c_0$ of NaClO$_3$ depends strongly on the temperature $T$ of the solution. Therefore, it is essential to carefully control the temperature of the sample.
The lab was temperature controlled to $20\pm 1$~C, the Olympus IX83 microscope was temperature controlled to $25\pm 0.1$~C by a Cube{\&}Box temperature controller from Life Imaging Services (www.lis.ch), a temperature regulation fluid was controlled by a Julabo refrigerated - heating circulator to $25\pm 0.01$~C and flowed through a Peltier element heat exchanger on the oil immersion objective (Zeiss antiflex 63X) and to one side of Peltier elements on the crystal growth chamber. The other side of the Peltier elements were in direct contact with the metal of the crystal growth chamber. Thermistors were placed in the heat exchanger of the objective and in the metal of the chamber to enable temperature regulation and measurement. The heat flow between the temperature regulation fluid and the chamber was regulated by PID controlled Peltier elements powered by in-house built current amplifiers. The PID, microscope, illumination and camera controls were programmed in Matlab and 
Micromanager.

\subsection*{Supersaturation of the solution}
The solubility $c_0(T)$ is the molar concentration $c$ of NaClO$_3$ in water in equilibrium with a NaClO$_3$ crystal. The solubility of aqueous NaClO$_3$ solutions depends strongly on the temperature $T$. In the temperature range 0-50$^\circ$C the dependence is linear and the relative change in solubility with respect to a reference temperature $T_0$ is $c_0(T)/c_0(T_0) - 1=(T-T_0)/\delta T$, where $\delta T=163$~K~\cite{Crump1995,Kerr1985}.
A crystal in the sealed chamber is allowed to equilibrate with the solution at temperature $T_0$. In order to determine $T_0$, the temperature is increased until the previously perfect cubic crystal starts to dissolve at its edges. Then the temperature is adjusted until neither growth nor dissolution at the roundish corners can be observed.  Then the temperature is changed by $\Delta T=T_0-T>0$ to achieve the desired supersaturation $\sigma(T)=c/c_0(T) - 1=c_0(T_0)/c_0(T) - 1\approx \Delta T/\delta T+(\Delta T/\delta T)^2$. The temperature is controlled with accuracy $\pm 0.01$~K, thus $\sigma$ is controlled with accuracy $\pm 10^{-4}$. The growth of the crystal will consume ions from the solution and change the concentration (and supersaturation) of the solution. For the frequent case of crystal sizes $L$ smaller than L=200~$\mu$m, the concentration of the solution in the chamber can be approximated to be constant. For larger crystals, i.e., crystals with edge lengths L$>200\mu$m, the 
concentration of the bulk solution is corrected by the consumption of material by the growing crystal.

\subsection*{Reflection Interference Contrast Microscopy}
The crystals are observed using reflection interference contrast microscopy (RICM), which is based on the interference of the light reflected by the sample with the light reflected by the glass surface the sample is placed on \cite{Curtis1964}. This technique is mainly used 
to examine the contacts between biological cells and glass surfaces. Several improvements to this technique have been made including numerical analysis, which considers a finite illumination aperture and a finite Numerical aperture\cite{Gingell1979, Gingell1982,Limozin2009a, Theodoly2010} the usage of the so called antiflex technique to improve the signal to noise ratio \cite{Ploem1975}, dual-wavelength reflection interference contrast microscopy \cite{Schilling2004} and the usage of a thin coating to shift the contact area away from the first minimum in intensity \cite{Radler1993}. RICM has mainly been developed and used for measurements of absolute fluid film thickness of soft and biological matter and has reached a measurement precision of 2-30~nm~\cite{Schilling2004,Limozin2009a,Theodoly2010,Contreras-Naranjo2013,Huerre2016}.
%

A sketch of the principle of crystal growth measurements using RICM is shown in Fig. 1. The intensity 
\begin{align}
I=2 \pi \int_0^{\theta_{max}} \sin(\theta) I_0(\theta) \gamma(\theta) r(\theta) d\theta
	\label{eq:Itot}
\end{align}
detected at a pixel of the detector results from the angular spectrum $I_0(\theta)$ of the illumination, the optical response function of the system $\gamma(\theta)$ and a interference based reflectance factor $r(\theta)$. Since we use an objective with a high numerical aperture, a large part of the light enters with a non negligible angle $\theta$ in respect to the optical axis. Rotational symmetry is used in equation \ref{eq:Itot}.
The reflectance factor
\begin{align}
	r(\theta) = R_{g,s}(\theta)+R_{s,c}(\theta')+2\sqrt{R_{g,s}(\theta)R_{s,c}(\theta')}\nonumber \\
	\cos\left(4 \pi n_s \phi(\theta, \theta') \frac{\zeta(r)}{\lambda} +\pi\right)
\end{align}
is a function of the local distance $\zeta(r)$ between crystal and glass interface, the refractive indices of the glass ($n_g=1.52$), the crystal ($n_c=1.515$) and the NaClO$_3$ solution ($n_s(c)=1.32415+0.136 \frac{c}{100g+c} $, where $c$ is the concentration in g per 100g H$_2$0~\cite{Bedarida1990}).
$\phi(\theta, \theta') = \frac{1}{cos(\theta')}-\frac{n_s}{n_g}\tan(\theta')\sin(\theta)\leq 1$ denotes a relative phase difference, which depends on the angles $\theta$ between the optical axis and the beam in glass and $\theta' = \arcsin\left(\frac{n_g}{n_s}\sin(\theta)\right)$ between the optical axis and the beam in solution.
The reflectances $R_{g,s}(\theta)$ of the glass-solution interface and $R_{s,c}(\theta')$ of the solution-crystal interface are determined by the Fresnel equations using the respective refractive indices \cite{hecht2016optics,Bedarida1990}.
We expect that the resulting intensity - distance relation in the region from contact to the first maximum is well represented by assuming a uniform illumination ($I_0(\theta)$ = const.) and system response ($\gamma(\theta)$ = const.
) in combination with an effective numerical aperture or maximal angle.
In order to obtain this effective numerical aperture, we imaged the interference intensity from a spherical lens (Thorlabs, LA1540, focal length: f=15mm, radius 7.7mm) in contact with the cover slip in a saturated NaClO$_3$ solution. The intensity-distance relation for the thus obtained calibration measurement is shown in Fig S1. 
We selected the effective numerical aperture to match the position of the first maximum determined by the calibration measurement, i.e., NA=1.23. 
Since the angular spectrum $I_0(\theta)$ of the measurement system does not depend on the reflectivity and thus the material this value is also valid for the crystal experiments. 
In order to reconstruct the distances $\zeta$ between the crystal and the cover slip below the first maximum of the interference, we use the min/max method by Limozin et al. \cite{Limozin2009a} in combination with the calibration described above. In the experiments, a ThorLabs SOLIS-525C LED with a centroid wavelength of 525~nm has been used in combination with a 16~bit Andor Zyla 5.5 sCMOS camera.

\subsection*{Measuring step front velocities}
The image sequences obtained in our experiments allow us to measure all the velocities of the step flow necessary to determine the complex step front dynamics. In order to determine the dependence of the velocity on the orientation of the two sub-layers, we selected areas at the boundary of the confined crystal interface (see regions a and b in Fig. 3) with the criteria that in the entire area both types of layer fronts are perpendicular to the outer border and not approaching it. This ensures that the component along the boundary is measured. To avoid inter step front effects, areas and time periods where selected in which one front is not catching up with the previous front. For a direct comparison of the different step velocity components, we chose two areas at adjacent boundaries of a single measurement. Thus, we guarantee the same supersaturation at both areas.

In the yellow and blue sub-regions in Fig.~3~D, (a and b) we have measured the step flow propagation velocities of the two half layers along the outer boundaries where the solution supersaturation is constant. In the selected areas, the pixel values were averaged in perpendicular direction to the border and filtered (Gaussian, standard deviation of kernel: $s=2$) in the parallel direction. The step front was detected by finding the peak in the differentiated curve.

\section*{Results}

\subsection*{Nucleation of molecular layers}
We found that on the (001) surface of NaClO$_3$ the minimum growth step height equals $z_0$=0.33~nm. This observation shows that NaClO$_3$ grows in an interlaced manner with two alternating monolayers, which differ in growth kinetics and which sum up to the thickness of one elementary cell height of $2z_0$=0.66~nm~\cite{Dickinson1921}. 

In about 95\% of our experiments the crystals had no dislocations and we did not observe any growth 
for a supersaturation $\sigma < 0.048$. Above that threshold new layers nucleate on the confined facet and form two-dimensional monolayer islands that propagate until they cover the facet (see Fig.~2~A and Movies S1-S3~\cite{MatMet}). On the length scale of our spatial resolution (300~nm) and time resolution (0.1~s) we observe that the {\em growth steps} at the edge of these monolayers {\em flow unimpeded} as if no solid was in direct contact with the growing crystal (see Movies S1-S5).
A layer of spacer particles of diameter 10-80~nm is dispersed between the glass coverslip
and the NaClO$_3$ crystal~\cite{Kohler2018} (see Fig.~S2~\cite{MatMet}), both to mimic a rough contact and to control the distance $\bar{\zeta}$.
Each time a new layer is added on the nanoconfined surface, the crystal surface is pushed back by the disjoining pressure and relaxes towards its equilibrium position (see Fig.~2B and Fig.~S3), raising the macroscopic crystal by one molecular layer. A few molecular layers of fluid therefore appear to be sufficient for growth steps to propagate  unaffected by the presence of spacer particles. The same effect of the disjoining pressure, but without spacer particles and with lower resolution, was observed for CaCO$_3$ crystals~\cite{Li2018_CGD}.

\begin{figure}
    \centering
    \includegraphics[width=7.5cm]{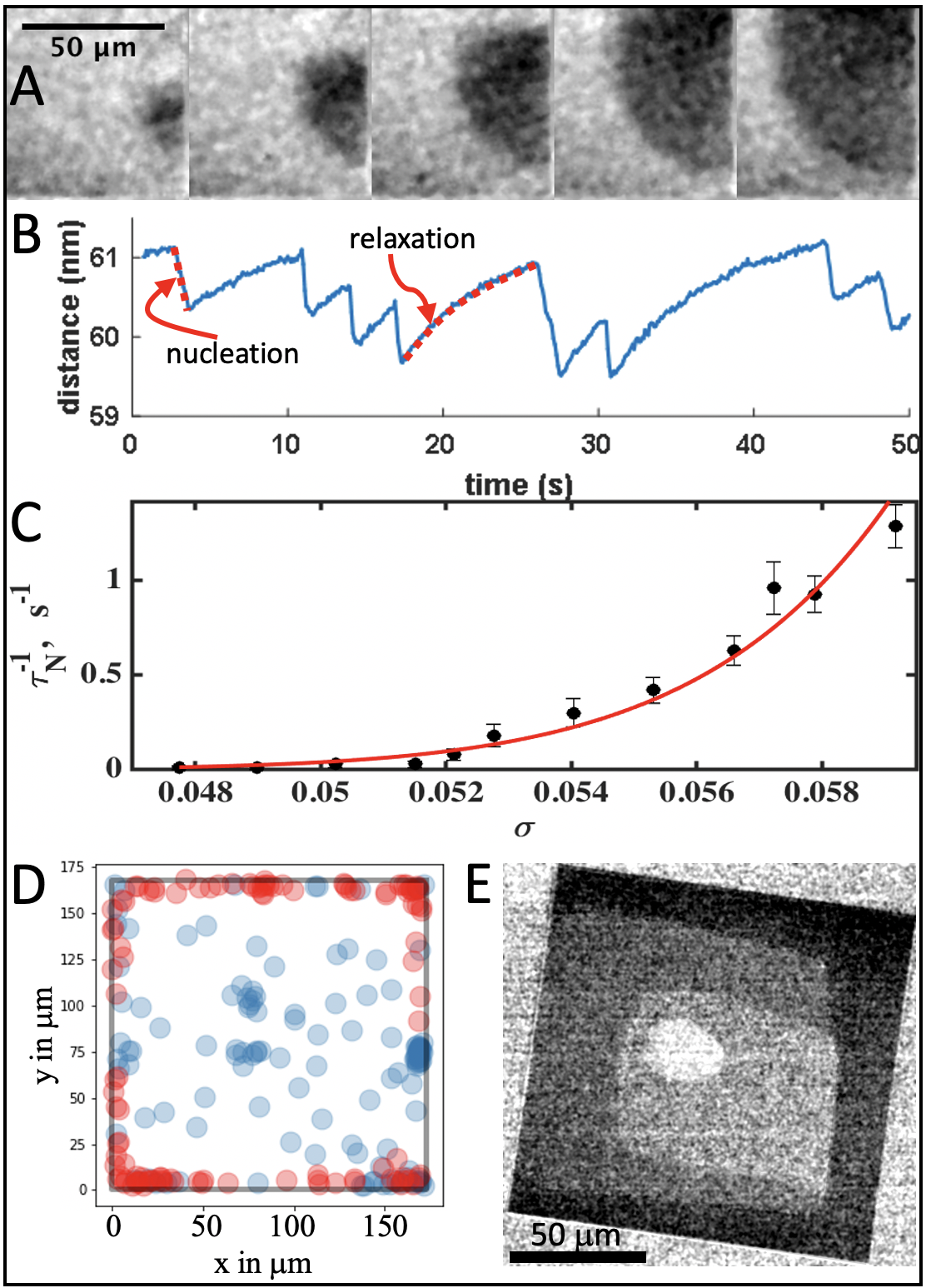}
    \caption{{\bf Nucleation of molecular layers.} {\bf A:} Average subtracted RICM images at 0.1~s interval of nucleation and spreading of a new layer 
at  $\sigma=0.051$.  
{\bf B:} Temporal evolution of mean distance, $\bar{\zeta}$,  for a small crystal (L=48~$\mu$m) showing
sudden nucleation events (steep negative slopes of one unit cell height, 0.66~nm) followed by a relaxation towards equilibrium distance. {\bf C:} Nucleation rate as function of supersaturation. The nucleation rate was determined by counting of new layers in distance--time curves like {\bf B}. The red line is a fit of nucleation theory to the data.
{\bf D:} Localization of nucleation at different supersaturations. Position of nucleation events at $\sigma \le 0.051$ (blue) showing random, homogeneous nucleation and a weak tendency of heterogeneous nucleation at three locations, and $\sigma > 0.051$ (red) showing strong localization at the edge.
{\bf E:} Enhanced RICM image from movie S2~\cite{MatMet} of crystal at $\bar{\zeta}=53$~nm, $\sigma=0.057$ showing four molecular layers with step fronts of height 0.33~nm advancing inwards. Nucleation at the crystal edges and faster step front
propagation in the vicinity of the crystal edge are caused by concentration gradients between the edges and the center.
}
    \label{fig:nucleation}
\end{figure}

By counting the new layers (rapid drops of 0.66~nm in $\bar{\zeta}$ in Fig.~2~B) we have measured the nucleation rate, $\tau_N^{-1}$, on the confined facet as function of supersaturation (see Fig.~2~C). The standard theory of nucleation predicts that
in the limit of small supersaturations $\sigma\ll 1$ corresponding to our experiments,
the nucleation rate per unit area is
$J=J_c(\sigma)\,{\rm e}^{-\sigma_c/\sigma}$~\cite{MatMet}.
The critical supersaturation $\sigma_c$ can be expressed in terms of physical quantities $\sigma_c=\pi\Gamma^2/4z_0^2$, where $z_0$ is the crystal step height and $\Gamma$ is the molecular line tension length scale~\cite{MatMet} and the critical nucleation rate $J_c$ depends only algebraically on $\sigma$.
Fitting the nucleation rate expression to experimental data  (see Fig.~2~C) yields  $\sigma_c=1.1\pm 0.1$ which leads to $\Gamma=0.40\pm 0.02$~nm.

We observe that at low nucleation rates nucleation may occur 
anywhere on the confined surface (see Fig.~2~D for $\sigma <0.051$ and Movie S1).
As supersaturation and nucleation rate increase
most molecular layers on the confined surface are nucleated close to the edge 
(see Fig.~2~D for $\sigma >0.051$).  
This abrupt change in nucleation localization is due to the depletion of ions in the confined fluid when a molecular layer grows. Diffusion does not have time to transport ions before the next nucleation event and a concentration gradient develops. Since nucleation depends exponentially on concentration it localizes at the outer edge.  The results in Fig.~2~D agree with the nucleation theory we have developed for confinement~\cite{MatMet}.

The number of monolayers of solid that can be formed by the ions in the fluid film is 
$\Theta_{eq}=\frac{\bar{\zeta}}{z_0}\frac{c_0}{c_s}(=\bar{\zeta}/1.2$~nm for NaClO$_3$),
where  $c_0$ and $c_s$ are the molar densities of the fluid and solid. 
The coverage, $\Theta_{eq}\sigma$, the number of monolayers of solid that can be formed from the {\em excess} of ions in the supersaturated liquid is
a relevant quantity for systems with two-dimensional mass transport such as surface diffusion
in molecular beam epitaxy (MBE)\cite{Saito1996}. Nanoconfined crystal growth is distinct from MBE in that the coverage can be larger than one.
For the experiment in Fig.~2C, $\bar{\zeta}\approx 48$nm and $\sigma=0.05$,
leading to $\Theta_{eq}\sigma\approx 2$. 
Due to the gradient in concentration the layers spread fast along the edges and slower inwards towards the centre of the confined surface as shown in Movie~S2. At higher supersaturations, $\sigma>0.058$, multiple steps may nucleate before the layers have spread across the entire confined surface. Once a new layer is nucleated and spreading, other step fronts closer to the centre slow further down because ions for growth are consumed by the outer step fronts and steps accumulate in a scenario similar to the bunching instability~\cite{Misbah2010}. 
Then the step bunch stops, leaving a ``cavity''~\cite{Kohler2018} (light grey region in Fig.~2~E, Fig.~S4 and Movie~S3) in the centre of the crystal surface.

\begin{figure*}
    \centering
    \includegraphics[width=16cm]{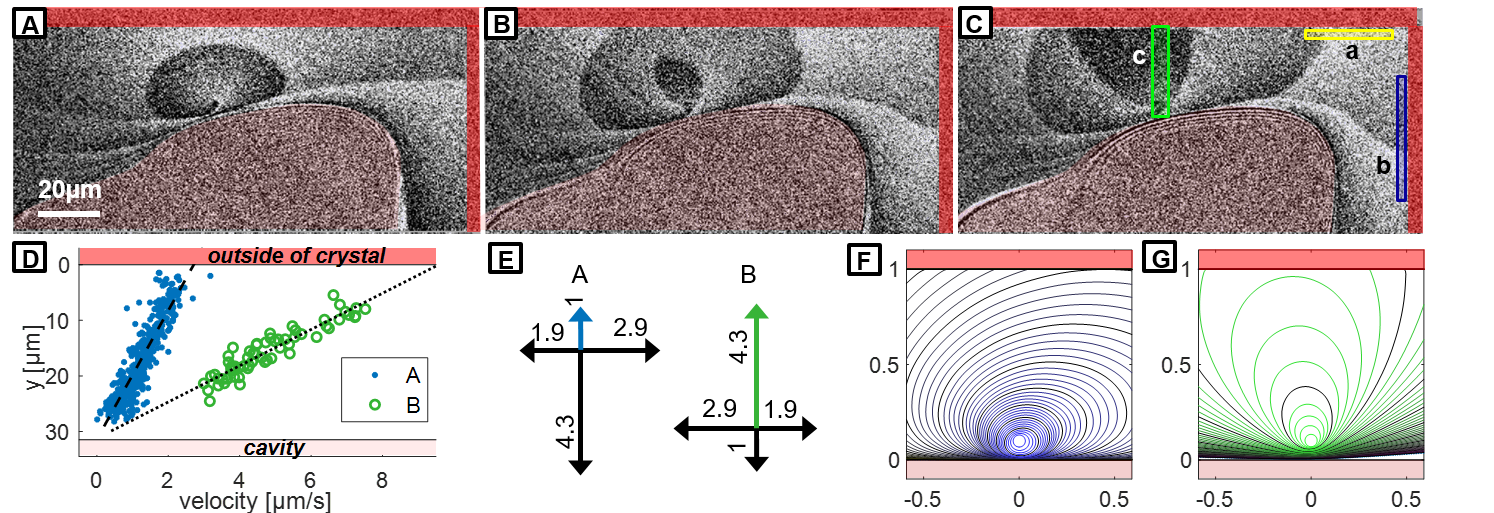}
    \caption{\bf{Spiral growth in nanoconfinement. A-C:}} Average subtracted RICM image series at a bulk supersaturation $\sigma=0.002$
with  3s intervals. The areas outside the crystal and inside the cavity are marked in respectively dark and light red colours. The oval regions of different intensity are molecular layers 0.33~nm in height each. The regions a and b in  {\bf{C}} have been selected for determination of the step flow velocities along the outer rim boundaries. The green region c in {\bf{C}} was selected to determine position (concentration) dependencies of step fronts shown in {\bf{D:}} position dependence of the step front propagation velocity in region c. The dashed and the dotted lines are linear fits to the data. {\bf{E:}} Kinetic anisotropy ratios of the two half layers A (blue) and B (green) corresponding to the slow and fast fronts shown in {\bf D}.  {\bf{F, G:}} Evolution of slow and fast step fronts simulated by simple forward step algorithm using a linear supersaturation gradient and kinetic anisotropy ratios displayed in {\bf{E}}. One observes close correspondence with the step front shapes in the 
images {\bf{A-C}}.
    \label{fig:spiral}
\end{figure*}

\subsection*{Spiral growth and step flow velocity}
In about 5\% of the observed crystals new molecular layers continuously originate from screw dislocations emerging at the surface like in Fig.~3 and Movie S5~\cite{MatMet}, even at supersaturations far below the 2D nucleation threshold.
The sharp jumps of intensity level are molecular step edges 0.33~nm in height each. 
The spiral atomic steps emerging from the dislocation are strongly
skewed towards the edge of the contact: while steps moving towards the edge are accelerated, those flowing towards the inner part  
where a cavity is present are slowed down. 
One observes that there are two alternating types of step edges,
one elongated in the vertical direction the other in the horizontal direction in the image, 
which correspond to the two half-unit cell interlaced layers.
Measuring the step flow velocity as function of orientation and position (see Fig.~3~D and~\cite{MatMet}) and assuming that supersaturation is proportional to the distance from the outer edge, 
the step velocity can be modeled as 
 $v(\sigma,\theta) = \alpha\sigma k(\theta)$, where 
$k(\theta)$ is an anisotropy ratio.
Simulations based on this law
shown in Fig.~3~F,G
are in close correspondence with experiments in Fig.~3~A-D
when $\alpha=1170\pm 110\mu$m/s with the anisotropy ratio reported in Fig.3.E.

\begin{table}
\begin{tabular}{l|l|l}
  & $v_A$, [$\mu$m/s] & $v_B$,  [$\mu$m/s]\\\hline
  region a: $\vec{x}$-direction & 5.4 & 8.2\\
  region b: $\vec{y}$-direction & 2.8 & 12.2
\end{tabular}
\caption{Step front velocities of the two interlaced layers A and B, measured at $\sigma=0.002$ on the crystal shown in Fig. 3.}
\end{table}

The front velocities in Table 1 were measured as described in Materials and Methods. Because there is a 180 degrees rotation of the chlorate orientation between two half layers (A and B) of a unit cell~\cite{Dickinson1921} the velocity field of layer B correspond to the one of layer A rotated by 180$^\circ$: $v_{A,-\vec{y}}=v_{B,\vec{y}}$ and $v_{A,-\vec{x}}=v_{B,\vec{x}}$. We define the kinetic anisotropy ratios $k_A(\theta)=v_A(\theta)/v_{A,\vec{y}}$ and $k_B(\theta)=v_B(\theta)/v_{B,-\vec{y}}$, where the angle $\theta$ is relative to the positive $\vec{y}$-direction. In Fig.~3~E we show the kinetic anisotropy ratios, $k(\theta)$, of the fronts. Note that the vertical and horizontal velocities are in principle for straight steps with a vanishing kink site density. The step velocity is normally an increasing function of the kink site density as well resulting in a function $k(\theta)$ whose principal directions are indicated in Fig.~3~E with a maximum velocity close to NW and SE respectively.




There is a difference in supersaturation from the outer rim (dark red), $\sigma=0.002$,
to the edge of the cavity (bright red), $\sigma\approx 0$. We have therefore measured the step front propagation velocity upwards in the image Fig.~3~C in the green region c. Fig.~3~D shows the measured velocities of the slow and fast fronts. The straight line fits demonstrate that the step front velocity is linear in vertical position, going to zero at the cavity and the maximum values (found in Table 1) at the outer edge.

In the very small range of supersaturations over the growth rim it is reasonable to assume that the supersaturation is linear in position, that is, the step velocity is linear in supersaturation:
\begin{equation}
  v(\sigma,\theta)=\alpha\sigma k(\theta), \label{eq:vel}
\end{equation}
where $\alpha=1170\pm 110\mu$m/s.

Using this relation we have simulated the slow and fast step front propagation from a dislocation source situated at the inner rim edge like in Fig.~3~A-C. The local step propagation velocity normal to the step front follows from (\ref{eq:vel}). The supersaturation $\sigma$ is assumed to be a linear function of the distance from the outer rim edge and constant in time in this stationary state. Each point of the front is shifted with every time step according to the local concentration and the orientation dependent growth kinetics as described by equation (\ref{eq:vel}). This emulates a part of an extended growth rim, where the inner (lower) part is connected to the cavity. For the orientation dependency of the step front kinetics the experimental values are used.
The fast and slow fronts have kinetic anisotropy ratios $k(\theta)$ as shown in Fig.~3~E.
The resulting slow and fast step fronts are shown in Fig.~3~H,I for every 10th simulation time step. One observes the close correspondence with the step fronts in Fig.~3~A-D. This shows that the simple assumptions of our simulations are valid for a crystal surface confined about 50~nm from an inert surface, with a small bulk supersaturation ($\sigma$=0.002)
and a growth spiral that ensures that step fronts pass with regular intervals (every $\approx$10~s).

\subsection*{Step flow instability}
When the distance between the crystal and the  substrate is small, we observe that the otherwise smooth step fronts
can develop protruding ``fingers'' (see Fig.~4 and Movie S4). The smooth, curved step
propagates along the facet edge and then inwards 
towards the facet center. At a distance $\approx 50\mu$m from the edge the
step destabilizes with an initial wavelength $\lambda\approx 12$~$\mu$m. 
The tips of the fingers advance at constant speed $55\pm 5\mu$m/s  in a direction of maximum kinetic anisotropy
whereas the slowest parts of the step front slow down as $v(t)=v(t_0)\sqrt{t_0/t}$, where $v(t_0=1.8s)=20\pm 3\mu$m/s is the velocity when the front destabilizes.

\begin{figure*}
    \centering
    \includegraphics[width=16cm]{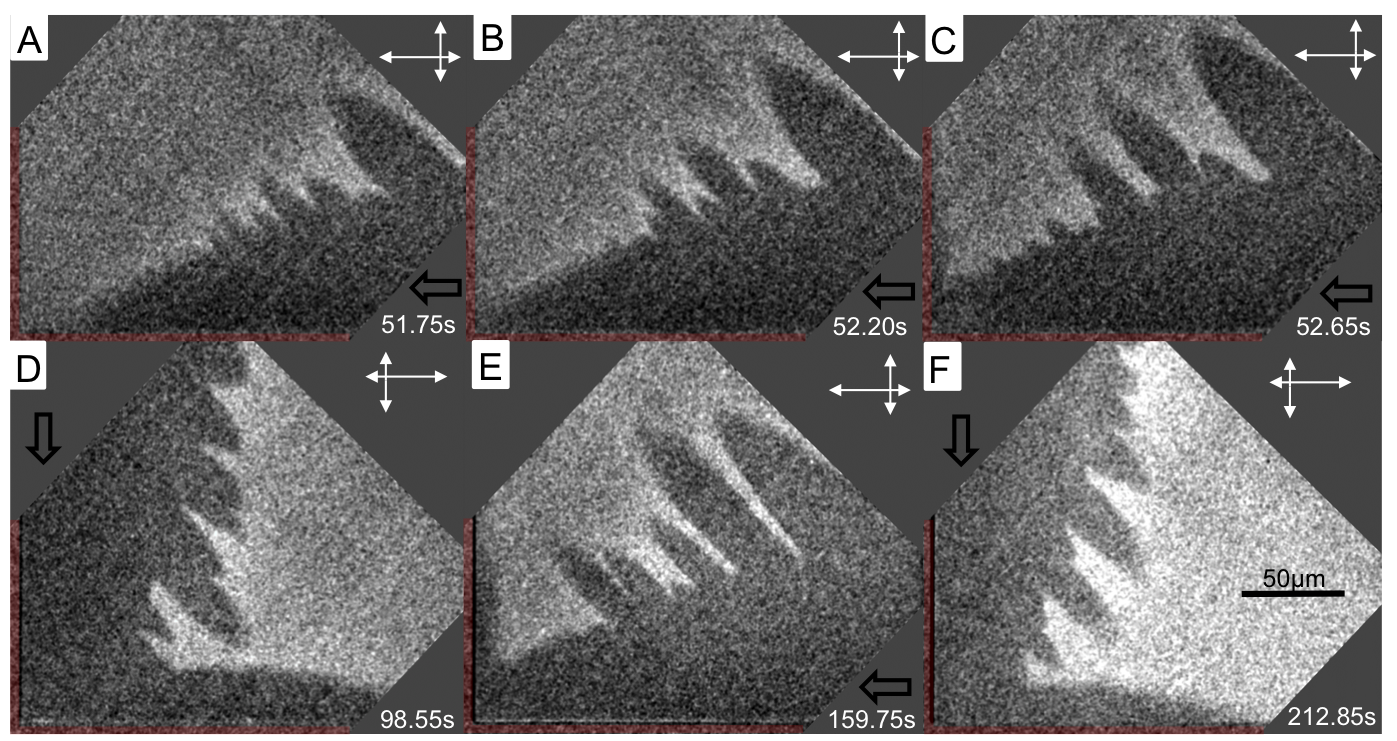}
    \caption{{\bf Step front instability.} Average subtracted RICM images of corner of a 
700$\times$700~$\mu$m$^2$ crystal
with $\sigma$=0.053 and $\bar{\zeta}$=22~nm.
The dark areas correspond to a smaller distance to the confining glass and thus to the newly formed layer showing instabilities at the front.
{\bf A-C:} Time lapse of the same step front at 0.45~s intervals {\bf C-F:}  Four consecutive step fronts nucleated at opposite sides with a time delay of 54$\pm$7~s.
The black arrow indicates the direction from which the layer originates. The orientation dependent growth kinetics of the respective layer is indicated by the white arrows. }
    \label{fig:instability}
\end{figure*}

We interpret this instability to be a variant of 
the Mullins-Sekerka instability 
which leads to complex growth shapes like dendrites and snowflakes~\cite{Saito1996,Ben-Jacob1993}:  
protuberances ahead of the step have a higher probability to catch 
randomly diffusing growth units (here ions), and thus
grow faster than the other
parts of the step.  The wavelength $\lambda =  2\pi \sqrt{3 \ell \Gamma\Theta_{eq}}$ emerging from the instability accounts for a competition between this destabilizing point effect and the stabilizing effect of line tension, 
where $\ell=D/v_{step}\approx 20\mu$m is the length scale of concentration gradients
associated to the motion of the step at velocity $v_\mathrm{step}$. 
From $\bar{\zeta}=20$~nm we obtain a coverage 
$\Theta_{eq}\sigma=0.9$ which agrees well with the fraction of the surface covered by the fingers. Using 
$\Gamma=0.40$~nm as obtained from 
the nucleation rates, we find $\lambda\approx 5\mu$m, which is
of the same order of magnitude as
the observed initial wavelength.  
Also, the velocity and radii of the finger tips are in agreement with existing 
theories for dendrite tips~\cite{MatMet}.
When 
$\bar\zeta$ is larger,
the coverage is larger than one, and as suggested from 
model simulations in the literature~\cite{Ben-Jacob1993,Saito1996} there is no instability as in Fig.2.

\subsection*{Confined growth morphologies}
\begin{figure}
    \centering
    \includegraphics[width=8cm]{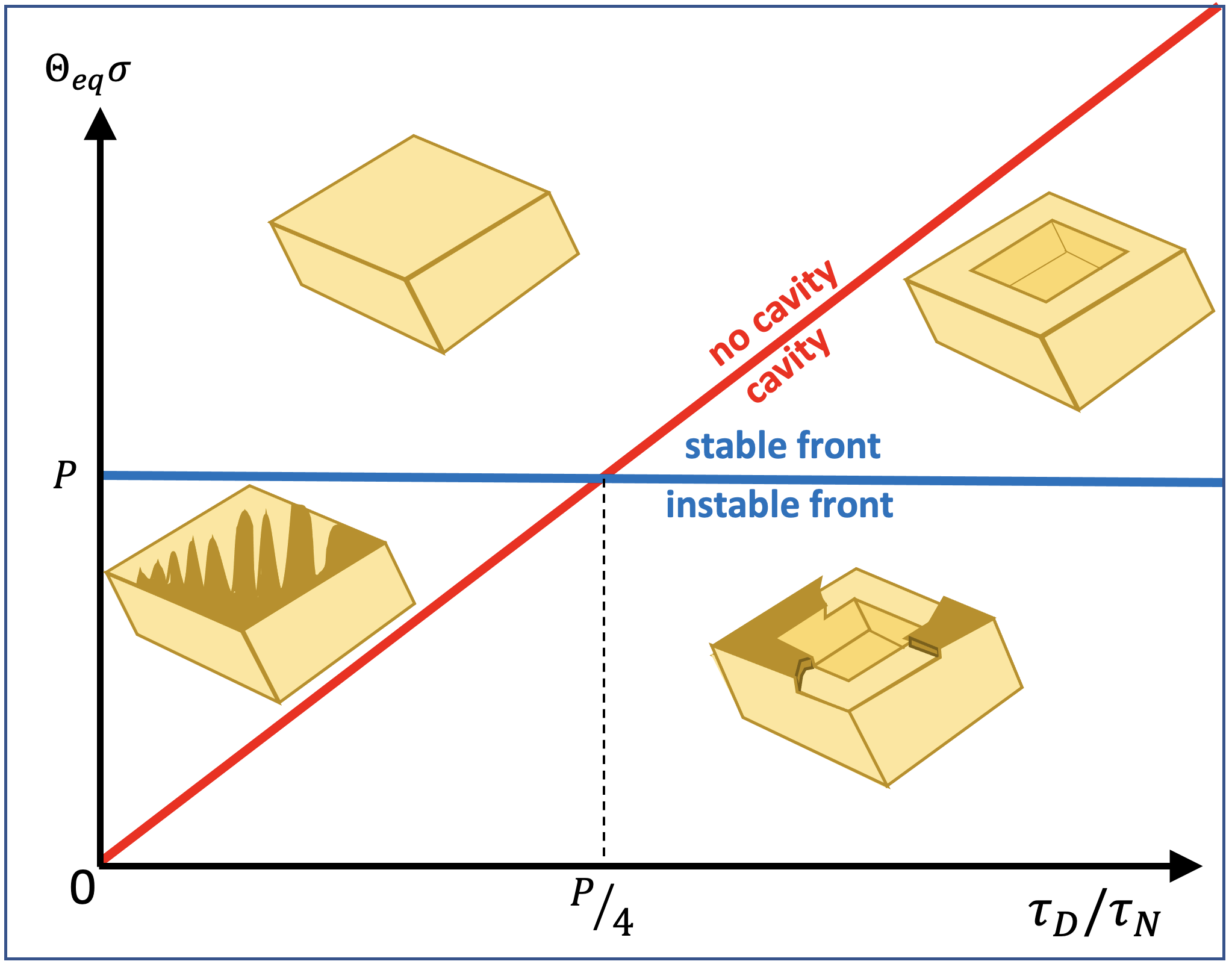}
    \caption{Nonequilibrium morphology diagram of confined growth morphologies. The stable step front morphologies (above the blue line) and the transition from no cavity to cavity (crossing the red line) have been observed both for NaClO$_3$ and CaCO$_3$ crystals. The finger-like instability to the lower left has only been observed on NaClO$_3$ and the growth rim roughening depicted to the lower right has been observed for both NaClO$_3$ and CaCO$_3$ crystals.}
    \label{fig:phasediagram}
\end{figure}
In order to 
reach a more general picture of
confined growth modes and morphologies we have also performed experiments on calcite crystals. The most important difference between NaClO$_3$ and CaCO$_3$ is that at room temperature the ratio of equilibrium solution concentration to solid concentration is $\frac{c_0}{c_s}\approx 0.3$ for NaClO$_3$ and $\frac{c_0}{c_s}\approx 2\cdot 10^{-5}$ for CaCO$_3$. We have measured the step flow velocity on confined calcite surfaces (see Figure~S6) and found that $\frac{v_s}{\sigma}\approx 10^{-8}$~m/s for CaCO$_3$ whereas $\frac{v_s}{\sigma}\approx 10^{-3}$~m/s for NaClO$_3$. The CaCO$_3$ step flow velocities now allow us to rationalize the previously observed transition from a smooth growth rim to a rough growth rim~\cite{Li2017b,Li2018_CGD}.

We will now reformulate the transition line (red line in Figure~\ref{fig:phasediagram}) from no cavity to cavity (measured both on NaClO$_3$~\cite{Kohler2018} and CaCO$_3$ crystals~\cite{Li2017b,Li2018_CGD}) on the confined surface as
$\Theta_{eq}\sigma=4\frac{\tau_D}{\tau_N}=\frac{L^2}{D}\frac{u_z}{z_0}$,
where $L$ is the size of the crystal or width of the growth rim, $u_z$ is the vertical growth rate and $\tau_D=L^2/D$.

The transition from a stable to an unstable front requires that the coverage is below 1 and that the step front outruns diffusion. In Figure~S5 one observes that the instability at coverage $\Theta_{eq}=0.9$ appears when the distance $L$ from the crystal edge is larger than the diffusion length $D/v_s\approx 50$~$\mu$m. For calcite on the other hand, the coverage is always far below 1. But since the requirement for advancing the step is to grow one monolayer, the balance between diffusion and step velocity must be multiplied by the minimum concentration, that is the coverage. The transition line for stable/unstable step front is therefore
$\Theta_{eq}\sigma=\frac{v_sL}{D}=P$, where $P$ is the Peclet number.
We can now test this prediction on the calcite smooth to rough rim transition. Using the step flow velocity $v_s=20-40$~nm/s at $\sigma=0.6$ we have two experiments to compare with. In Li et al~\cite{Li2017b} Figure~7 one observes that a calcite crystal with $\zeta\approx 40$~nm goes through the transition as the crystal grows from $L=4$~$\mu$m to $L=8$~$\mu$m, which corresponds to $\Theta_{eq}\sigma/P$ going from 1 to 0.5, thus crossing the blue line in Figure~\ref{fig:phasediagram}. In Li et al~\cite{Li2021} the confinement is changed from $\bar{\zeta}=600$~nm to $\bar{\zeta}=40$~nm, triggering a rough growth rim. The change in $\bar{\zeta}$ changes $\Theta_{eq}\sigma/P$ from 3 to 0.2, once again moving the confined surface across the blue line in Figure~\ref{fig:phasediagram}. It must be remarked that the nature of the instabilities are not the same for the fingering instability on NaClO$_3$ and the rim roughening of CaCO$_3$. The first one, studied in detail above is in a single step regime, while on the calcite growth rim there are multiple steps. The latter transition is probably due to step bunching but detailed, high resolution experiments will be necessary to clarify this. Both instabilities are triggered by competition between step flow growth and diffusion in a confined environment and the pertinent transition line is therefore coincident.


\section*{Conclusions}
We have established that crystal growth steps can flow freely in confined interfaces unhampered by contact heterogeneities. The inherent transport restrictions in nanoconfinement cause a distinctive growth mode characterized by two-dimensional transport of ions in the liquid film with gradients between the edge and the center of the contact that control the morphology and growth rate of nanoconfined crystal surfaces. 
These gradients are constitutive and therefore unavoidable in nanoconfined growth and free motion of molecular steps controlled by these gradients defines a nanoconfined growth mode that is different from known growth regimes.
Some specific features associated to nanoconfined growth are the localization of nucleation along the contact edge, strongly skewed spirals, and 
control of the stability of molecular steps by the distance to the substrate.

Our observations demonstrate that high resolution optical measurements of nanoconfined surfaces paves the way for \textit{in situ} observation  and identification of specific crystal growth regimes of relevance for biomineralization, templated crystal growth for advanced materials and crystallization pressure.

The general growth morphology diagram, with dimensionless parameters that can be measured or estimated for any confined crystal growing from solution, can be applied to materials design, conservation, biomineralization and geological crystal growth.

\acknowledgments
This project has received funding from the European
Union’s Horizon 2020 research and innovation programme
under the Marie Sklodowska-Curie Grant Agreement
No. 642976 (ITN NanoHeal) and from the Norwegian
Research Council Grant No. 222386. We thank Henrik Sveinsson for preparing MD configurations for Figure S2.

\bibliography{stepFlowBib}

\end{document}


\title{Supporting Information for Crystal growth in confinement} 
\author{Felix Kohler}
\affiliation{The NJORD Centre, Dept of Physics, University of Oslo, P.O.box 1048 Blindern, 0316 Oslo, Norway}
\author{Olivier Pierre-Louis} 
\affiliation{Institut Lumi\`{e}re Mati\`{e}re, UMR5306 Universit\'{e} Lyon 1 - CNRS, 69622 Villeurbanne, France}
\author{Dag Kristian Dysthe}
\affiliation{The NJORD Centre, Dept of Physics, University of Oslo, P.O.box 1048 Blindern, 0316 Oslo, Norway}

\begin{abstract}
 This PDF file includes:

 Captions for Movies S1 to S6

 Materials and Methods

Supplementary Text

 Figs. S1 to S7

Other Supplementary Materials for this manuscript include the following: 

Movies S1 to S6

\end{abstract}
\maketitle

\section*{Movies}

\subsection*{Movie S1}

Single molecular layers nucleated and propagating over crystal surface at low supersaturation ($\sigma=0.051$) and low nucleation rate. Timelapse movie of average subtracted RICM images at 0.1~s interval.  The dark areas correspond to a smaller distance to the confining glass and thus to the newly formed single molecular layer 0.33~nm thick. The four nucleation events originate at different locations. The crystal is $160\times160\mu$m.

\subsection*{Movie S2}

Single atomic layers nucleated and propagating over crystal surface creating concentration gradient and tending towards cavity formation.  Timelapse movie of average subtracted RICM images at 0.1~s interval.  The dark areas correspond to a smaller distance to the confining glass and thus to the newly formed single molecular layer 0.33~nm thick. Same crystal as in Movie S1 and Fig. S4 with $\sigma=0.055$. Due to higher nucleation rate than in Movie S1 the diffusion of ions does not replenish ion concentration at the centre resulting in a concentration gradient from edge to centre. Due to the concentration gradient nucleation centres are located at the crystal edge and the molecular layer step flow is slower towards the centre than along the edges. Fluctuations in nucleation frequency causes fluctuations in concentration gradient and stability of cavity.

\subsection*{Movie S3}

Fluctuations in nucleation and cavity formation. Timelapse movie of average subtracted RICM images at 1~m interval. Same crystal as in Movies S1 and S2 and Fig. S4 A and B at $\sigma=0.06$. The crystal size increases during the movie thus increasing the diffusion time and the criterium for cavity formation is fulfilled: $\Theta_{eq}\sigma\tau_N/\tau_D<1$. Fluctuations in nucleation frequency causes fluctuations in the stability of the cavity. A cross-section versus time of this movie is displayed in Fig S4.

\subsection*{Movie S4}

Step front instability. Timelapse movie of average subtracted RICM images of corner of a 700x700~$\mu$m$^2$ large crystal with $\sigma$=0.053 and $\bar{\zeta}$=22~nm.
  The dark areas correspond to a smaller distance to the confining glass and thus to the newly formed single molecular layer 0.33~nm thick, showing instabilities at the front. The four instable fronts propagate in the fast direction and display fingers that do not cover the entire surface move at constant velocity while a slow (and slowing down), diffusion controlled front fills the layer between the fingers. Some images from this movie are shown in Figs. 4 and S5.

\subsection*{Movie S5}

Spiral growth on a nanoconfined growth rim. Timelapse movie of average subtracted RICM images at 1~s interval.  The oval regions of different intensity are molecular layers 0.33~nm in height each.  Images in this movie are also displayed in Fig.~3. The top and right hand side of the movie is the crystal edge and trhe grey area at bottom left is the cavity inside the growth rim. One observes that every second layer has different propagation velocity and shape due to the different (rotated 180 degrees) kinetic anisotropy ratios of the two half layers constituting a unit cell of the crystal.

\subsection*{Movie S6}

Corner of calcite crystal 30 nm above confining glass surface in water with 0.8 mM CaCO$_3$ concentration ($\sigma=0.6$) imaged once every minute for 90 minutes. The waves moving from bottom left to top right are interference contrast of growth steps propagating on the confined surface.

\section*{Estimation of disjoining pressure in crystal-spacer contacts}
\begin{figure}
    \centering
    \includegraphics[width=1.00\columnwidth]{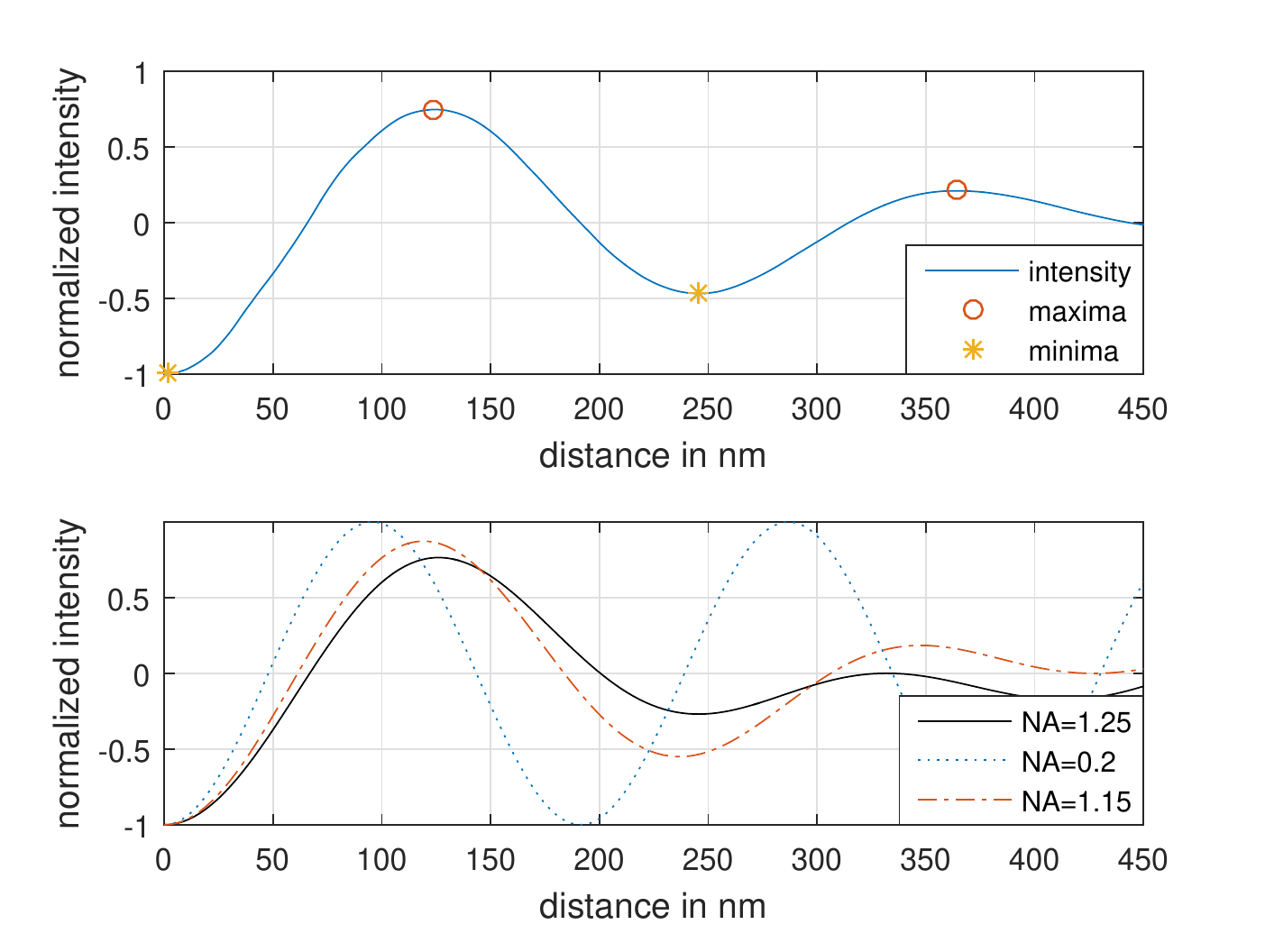}
    \caption{Intensity of the interference term as a function of the distance d to the glass a) Reference measurement performed with a spherical lens ($r=7.7mm$) b) Intensity as a function of d for different effective numerical apertures calculated using the equations in the text.}
    \label{fig:S1}
\end{figure}

When two surfaces approach each other the liquid film between the surfaces can support a normal stress without being squeezed out~\cite{Israelachvili2011}. This is termed the {\em disjoining pressure} of the fluid film on the solid surfaces. 
{ In the nanometer range where the disjoining pressure is probed 
in our experiments, van der Waals forces are negligible. The electric double layer contribution dominates and }
scales with the Debye length and at small fluid film thickness (1-5 molecular layers) there are so-called steric repulsive forces that increase rapidly with decreasing distance $z$. In order to squeeze out the last 2-3 molecular layers of water, pressures of the order 1~GPa are required~\cite{Israelachvili2011}. Thus at high ion concentration one expects the fluid film thickness $z$ to be in the range 0.8-1.6~nm for all pressures in the range 0.1~MPa to 1~GPa.
Even though the liquid can support a normal pressure it still behaves like a fluid in the sense that diffusion is almost as fast as in bulk until the liquid film thickness $z$ is reduced to 2 molecular layers~\cite{Mutisya2017}.

\begin{figure*}
    \centering
    \includegraphics[width=12cm]{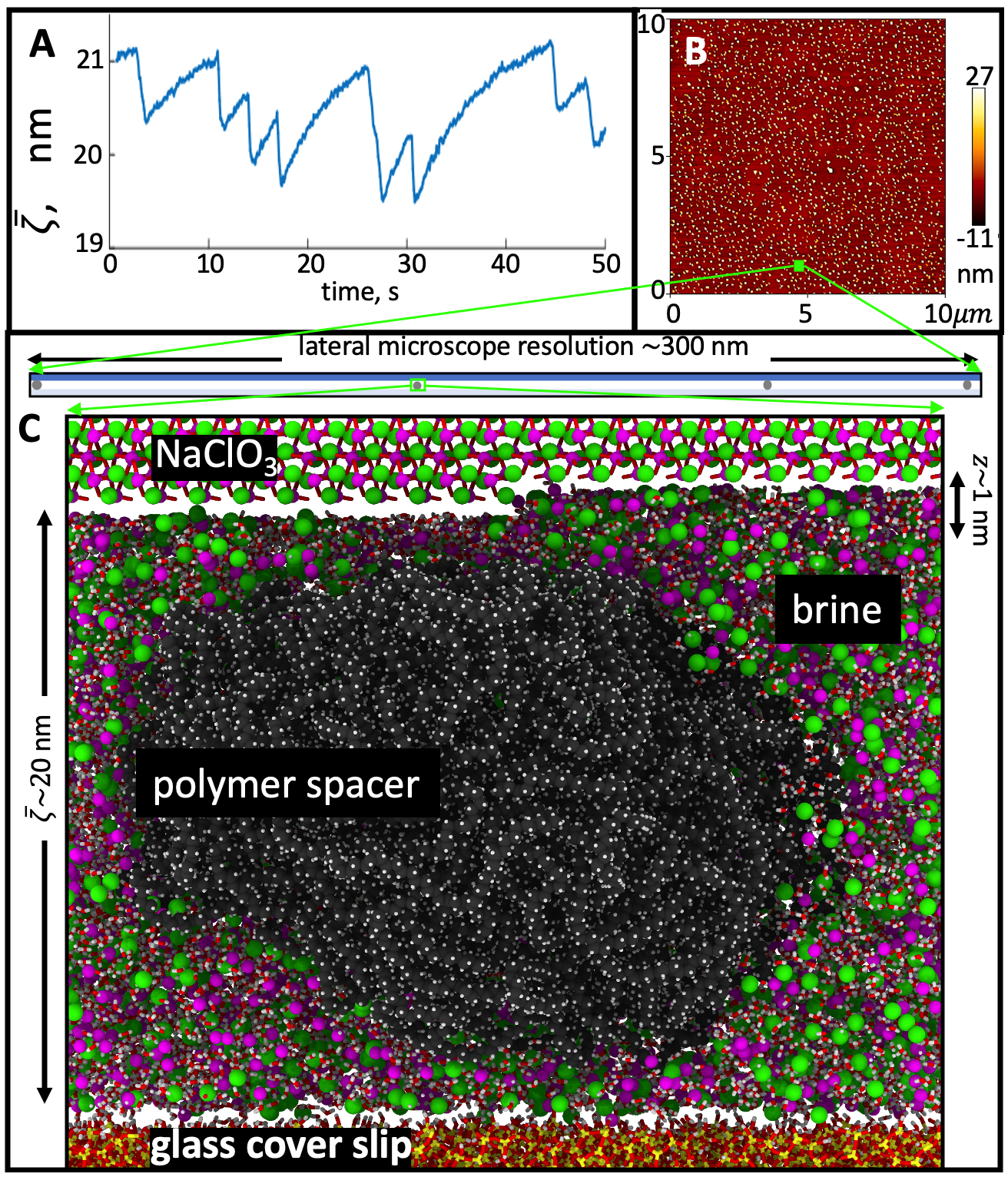}
    \caption{Step flow through contacts. {\bf A} The average gap distance, $\bar{\zeta}$ between the crystal and the glass support as function of time. The rapid drops by 0.66~nm in $\bar{\zeta}$ correspond to two molecular layers (one unit cell thickness) growing on the crystal in the confined region by step flow. After the gap has been reduced it slowly relaxes towards the equilibrium distance again. {\bf B} Atomic force microscope (AFM) image of glass surface before mounting of cell. The glass support is evenly covered by spacers of radius $r\approx 20$~nm with a spacer density of $\approx 18 \mu$m$^{-2}$. {\bf C}  Vertical cut of crystal, spacers and glass illustrating the crystal step edge propagating through the thin ($\approx$1~nm) liquid film that supports the weight of the crystal. The image illustrates the correct relation between 10~nm spacer size, 0.33~nm step height and $\approx$1~nm fluid film thickness in the contact. The densities of water and ions in crystal and solution are also to scale. When two steps have passed, the liquid layer has become 0.66~nm thinner and the disjoining force in the liquid layer increased. This increased force pushes the crystal back towards equilibrium as observed in {\bf A}.}
    \label{fig:S2}
\end{figure*}

In our experiments the growing crystals are resting on small spacers that keep the distance $\zeta$ between the growing crystal and the supporting glass surface between 10 and 130~nm. If there were no spacers the { equilibrium} distance would be of the order 1~nm. This would make the optical contrast of the RICM technique too low to accurately measure height differences and the rate of ion transport by diffusion would be a factor 30-1000 lower, thus halting the process we wish to study. Figure~S2 shows the distribution and height of spacers measured by AFM, the change of $\zeta$ when a molecular layer is grown and the relaxation back to the equilibrium distance and an illustration of the step flow through the contact between a spacer and the crystal. The relaxation back to the equilibrium distance demonstrates clearly that the crystal is supported by the disjoining pressure in the fluid film in the contacts with the spacers.

The contact (disjoining) pressure can be estimated by noting that a cap of about 1~nm height of the spacer will contribute as support for the fluid film. Thus for an average spacer size of assumed spherical shape of diameter 20~nm the supporting area is $A_{cont}=\pi(20\sin\arccos 19/20)^2\approx 10^{-16}$~m$^2$. The NaClO$_3$ crystals of linear size $L$ are typically half as high as wide, thus their volume are $V=L^3/2$. The density difference of the crystal and the saturated solution is $\Delta \rho=$1100~kg/m$^3$. 
The equilibrium pressure in the contact can then be estimated as
\begin{equation}
  p_0=\frac{\Delta\rho g V}{2N_{cont}A_{cont}}= \frac{1}{2}\Delta\rho Lg\alpha = 5500 L\alpha,
  \label{eq:p0}
\end{equation}
where $\alpha=A_{cryst}/N_{cont}A_{cont}$ is the ratio of crystal area to contact area. This { ratio} can be estimated from Figure~S2~B where $A_{cryst}=(10\mu$m$)^2=10^{-10}$~m$^2$ and there are 1848 spacer particles. The real, unknown number of contacts $3<N_{cont}<1800$ depends on the size distribution and stiffness of the spacer particles. The area ratio is then $\alpha=A_{cryst}/N_{cont}A_{cont}\approx 10^6/N_{cont}\approx 550 - 3.3\cdot 10^5$. It follows that for a typical crystal size used in our experiments, $L=100 \mu$m the disjoining pressure in the spacer-crystal contacts are in the range $300-2\cdot 10^5$~Pa. This indicates that even if the crystal is supported by only 3 spacer particles the contact pressure $p_0$ is only about one atmosphere, well below the pressure necessary to squeeze the fluid film thickness to only 1-2 molecular layers.

\begin{figure*}
    \centering
    \includegraphics[width=12cm]{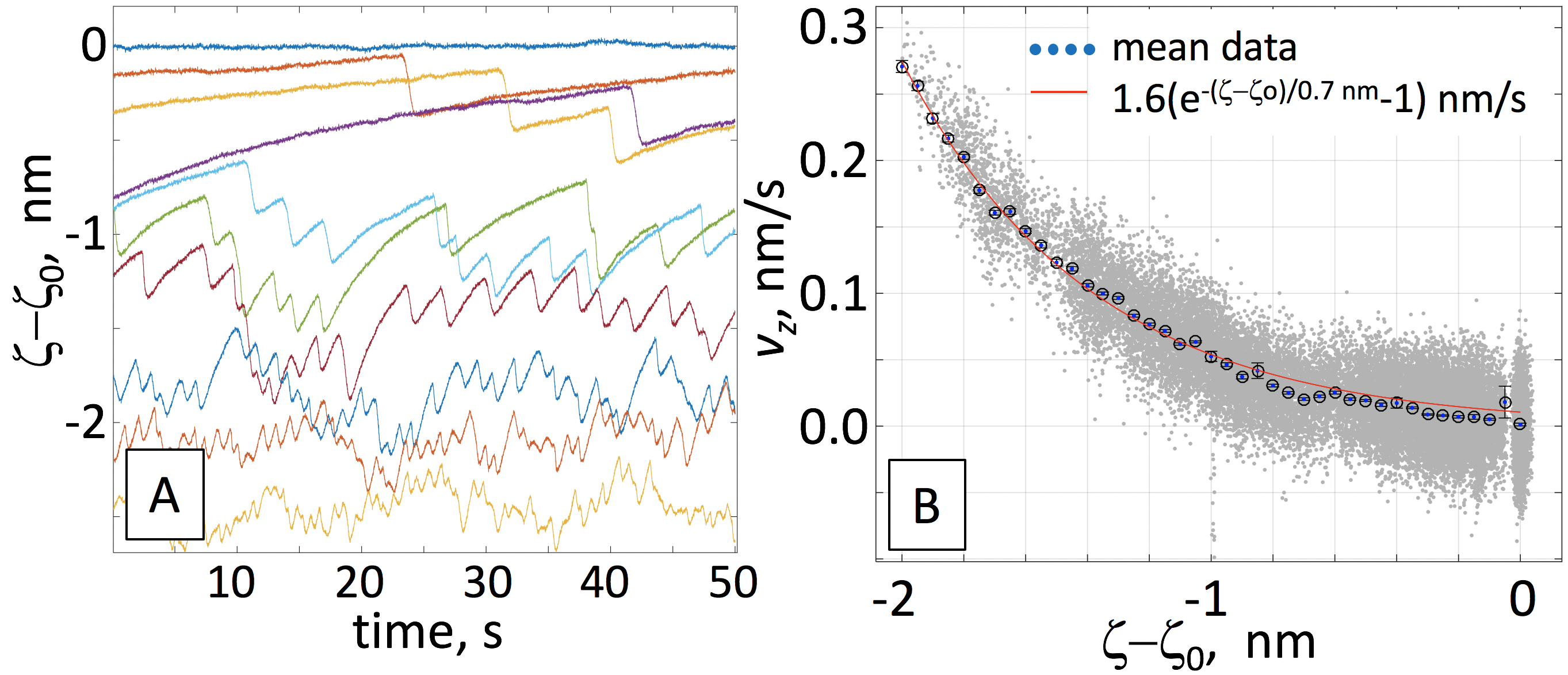}
    \caption{Relaxation of fluid film distance due to disjoining pressure. {\bf Left:} Mean fluid film thickness minus equilibrium value $\zeta-\zeta_0$ for many different nucleation rates. Growth of a new layer reduces the fluid film thickness by 0.66~nm. Excess disjoining pressure $\Delta p=p-p_0$, where $p_0$ is the equilibrium pressure pushes the crystal back towards the equilibrium pressure. Viscous drag of fluid flowing in the fluid film to allow the relaxation opposes the vertical motion.  {\bf Right:} Vertical velocity of crystal as function of $\zeta-\zeta_0$ from derivative of positions in left subfigure.}
    \label{fig:S3}
\end{figure*}

We may use the relaxation of the fluid film thickness $z$
{ between the top of particles and the crystal}  
towards the equilibrium distance $z_0$ to probe the disjoining pressure further. Growth of a new layer reduces the fluid film thickness by 0.66~nm. Excess disjoining pressure $\Delta p=p-p_0$, pushes the crystal back towards the equilibrium distance,
{ while} 
viscous drag of fluid flowing between the crystal and the support opposes the vertical motion. We may assume that the disjoining pressure, $p$, at these distances is an exponential function of distance:
\begin{equation}
\Delta p=p-p_0=p_0(e^{-(z-z_0)/\lambda_D}-1),
\end{equation}
where $\lambda_D$ is the Debye length. 
{
Since the crystal moves as a whole, the change of the average distance $\zeta$ between the crystal and the glass substrate is equal to the change in the film thickness at the top of particles: $\zeta-\zeta_0=z-z_0$.}
A circular plate of radius $R$ at distance $\zeta$ from another surface that is subjected to a disjoining pressure $\Delta p$ 
{ at the contacts} will have the vertical velocity
\begin{eqnarray}
v_z&=&\frac{N_{cont}A_{cont}\zeta_0^3}{3\pi\eta R^4}\Delta p\\
&=&\frac{2\zeta_0^3\Delta\rho g}{3\pi \eta R} (e^{-(\zeta-\zeta_0)/\lambda_D}-1)\\
&=&v_{z,0}(e^{-(\zeta-\zeta_0)/\lambda_D}-1),
\end{eqnarray}
{ where $\eta$ is the viscosity, and we have used the approximation $\zeta\approx\zeta_0$
in the prefactor.}
 The crystal in the experiment shown in Fig.~S3 had $R\approx 50 \mu$m and $\zeta_0\approx 100$~nm. Using $\eta\approx 10^{-3}$~Pa~s the predicted velocity is $v_{z,0}\approx 0.2$~nm/s. For a monovalent salt at 7 molar the theoretical Debye length is 0.13~nm and the high concentration decay length has recently found to be 3~nm~\cite{Smith2016}. 

In Fig.~S3 we show the displacement curves at different nucleation rates and the agglomerated velocity - distance curve with the fit yielding the model parameters $v_{z,0}=1.6$~nm/s and $\lambda_D=0.7$~nm. This is reasonably close to the $v_{z,0}$ predicted from viscous drag and the decay length is between the two estimates given above. Based on the fitted Debye length we expect the crystal and weight supporting grains to be separated by only 3-4 layers of adsorbed solution~\cite{Israelachvili2011}. If the fluid film thickness approached 1 molecular crystal layer one would expect that the step flow could be pinned in the spacer-crystal contacts. In only very few experiments have we observed such pinning of the crystal growth step front by spacer particles.

\subsection*{Nucleation}
\begin{figure}
    \centering
    \includegraphics[width=8cm]{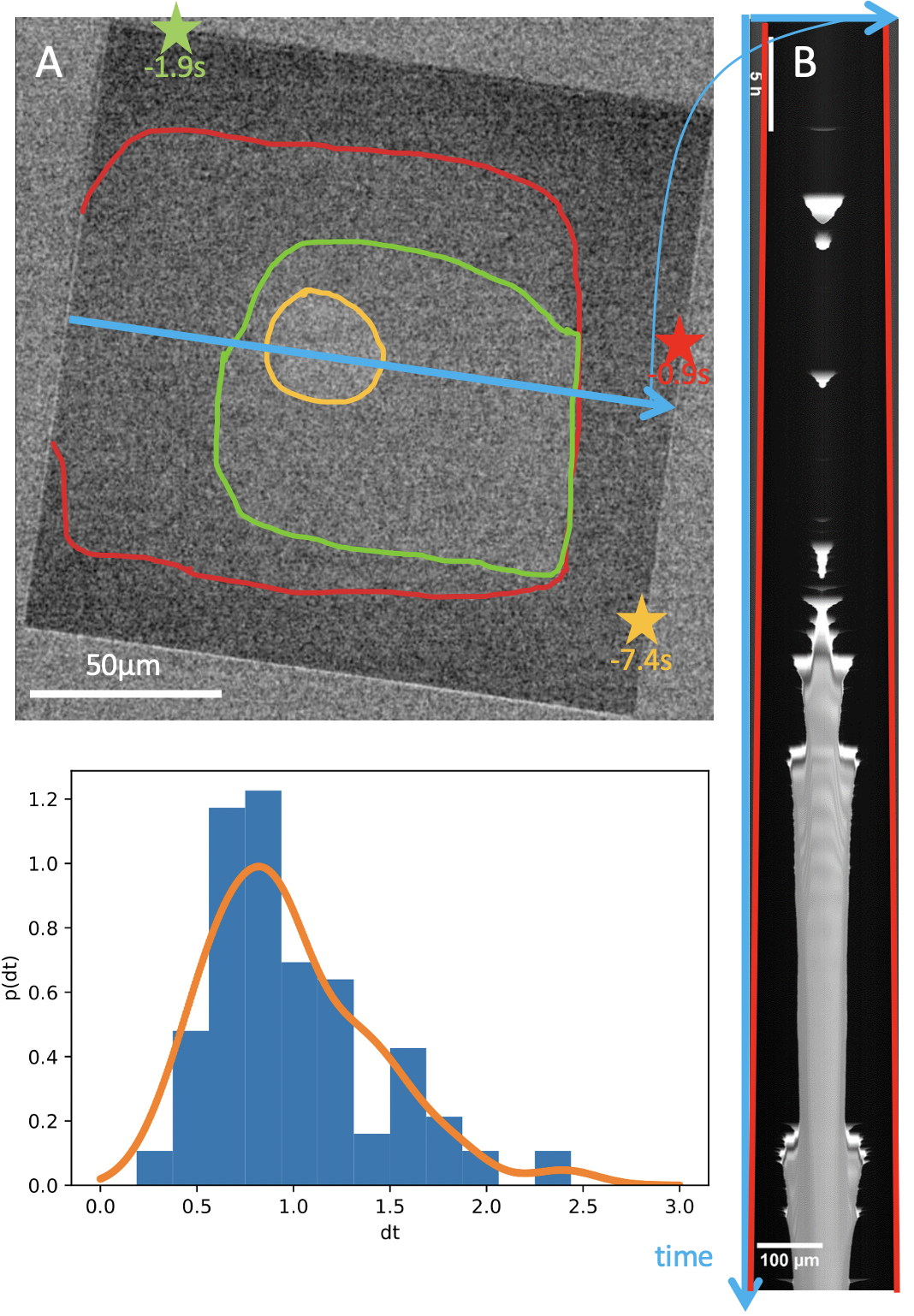}
    \caption{Fluctuations in nucleation and cavity formation. The high speed and high vertical resolution of our measurement technique allows us to measure the distribution of the nucleation rates as shown in subfigure {\bf A}. Subfigure {\bf B} is the intensity along the blue arrow in subfigure {\bf A} during the entire Movie S3. The crystal size (indicated by red lines) increases continuously with time and finally exceeds a threshold value where the cavity is stable and grows with fluctuations in size similar to the initial fluctuations. The width of the rim (black) outside the cavity (bright) is smaller than half the crystal size during the fluctuating cavity period. This corresponds to an unstable regime between two ``phases'' (no cavity and cavity) of the nanoconfined system. The fluctuation between cavity and no cavity and fluctuation of the cavity size is due to the fluctuation in time between nucleation of new layers, dt, displayed in subfigure {\bf C}.}
    \label{fig:S4}
\end{figure}

\subsubsection*{Standard nucleation theory}
The standard theory of nucleation is reported in many books and lecturenotes on crystal growth~\cite{Saito1996}. In this theory, the free energy of a monolayer island is composed of two contributions. The first one is the chemical potential
gain $\Delta\mu$ caused by the crystallization of the ions
when the ionic concentration in the liquid exceeds the solubility. 
The second contribution is the free energy cost of the formation
of the atomic step surrounding the monolayer island.
Since the first contribution is proportional to the area of the island
while the second is proportional to the perimeter, the 
second term always wins for small islands. As a consequence, 
there is a free energy barrier that must be surmounted
in order to form the two dimensional layer.
The passage over this barrier is triggered  by random thermal fluctuations.
Once the size of the island is larger than the critical size
corresponding to the energy barrier,
the islands grow irreversibly to decrease their total free energy.

We assume small supersaturations $\sigma\ll 1$, so that the chemical
potential reads
\begin{eqnarray}
\Delta\mu=nk_BT\ln(1+\sigma)\approx nk_BT\sigma
\end{eqnarray}
where the factor $n=2$ comes from the presence of two ions in the liquid
for one molecule in the solid. If the solution is not ideal,
the chemical potential is still proportional to $\sigma$
for $\sigma\ll 1$,
but the factor $n$ can be affected when taking into account the
activities of relevant species in the fluid.

Under the assumption of isotropic step properties, the rate of formation of new monolayer islands per unit 
facet area in the presence of a supersaturation $\sigma$ reads
\begin{eqnarray}
\label{eq:nucl_theory}
J=J_c(\sigma)\,{\rm e}^{-\sigma_c/\sigma}
\end{eqnarray}
where
\begin{eqnarray}
J_c(\sigma)&=& \frac{\rho_s \bar\alpha}{\Omega_2^{1/2}}(n\sigma)^{1/2}
\\
\sigma_c&=&\frac{\pi\Gamma^2}{\Omega_2 n}
\end{eqnarray}
We have defined the concentration $\rho_s$ (number per unit area) of initial seeds for nucleation, the orientational average of the kinetic coefficient $\bar\alpha=\alpha\langle k(\theta)\rangle$ discussed in the main text, the molecular area $\Omega_2$ in a monolayer 
(in the case of NaClO$_3$, we have $\Omega_2=2z_0^2$ where $z_0$ is the step height)
and $\Gamma=\Omega_2 \frac{\gamma}{k_BT}$, where $\gamma$ is the step free energy per unit length, and $k_B$ is the Boltzmann constant.   

The two main conditions of validity of (\ref{eq:nucl_theory})
are: (i) $\sigma\ll\sigma_c$, which ensures that the nucleation energy barrier
is larger than $k_BT$, and (ii) nothing special
occurs at the molecular scale when the nucleus is composed
of a small number of molecules, i.e., no extremely stable
intermediate molecular configuration and no extremely slow process
at the molecular scale.

\subsubsection*{Localization of nucleation}

If the supersaturation is the same everywhere on the facet,
then the total nucleation rate is simply $JL^2$ 
where $L^2$ is the area of a square facet of lateral extent $L$.
Assuming that the change of the facet size $L$
is negligible between two nucleation events,
as seen in experiments, the nucleation
time $\tau_N=1/(JL^2)$ given by the expression (\ref{eq:nucl_theory}) can be fitted to the data (see Figure 2)  using two free parameters:
$\sigma_c$ and $\rho_s$. We then find 
a surface coverage of initial seeds
for nucleation $\Omega_2\rho_s\approx 2.5\times 10^{-8}$
where $\Omega_2=2z_0^2$ is the area occupied
by one molecule in the monolayer.
In addition, we obtain $\sigma_c= 1.1\pm 0.1$,
leading to $\Gamma=0.40\pm 0.02$nm.

However, in general the supersaturation is not homogeneous on the 
confined facet. Along the periphery of the facet, the supersaturation
is expected to be constant due to continuous exchange
of mass with the neighboring bulk phase. 
However, somewhere on the facet, the supersaturation 
can be lower due to the growth of some step that have consumed
part of the ions in the liquid film. Such a depletion survives
even after the passage of a step during a time of the
order of $\tau_{D}=L^2/(4D)$ where $L$ is the lateral size of the facet.
Using $D=0.6\times 10^{-9}$m$^2.$s$^{-1}$, 
and $L=175\mu$m (as in Figure 2D),
we obtain $\tau_{D}\approx 10$s. This
timescale is of the same order as the
time between two nucleation events (from $1$ to $100$s)
in the regime where there is never more than
one step on the confined facet.
The absence of clear separation between these
two timescales prevents a quantitative prediction
of the supersaturation profile, which controls
the localization of nucleation close to the edge.
However, a generic analysis presented below 
catches the essence of the localization 
of nucleation.

In order to investigate the influence of 
supersaturation gradients on the localisation of
nucleation events, we consider a simplified 
one-dimensional geometry, with a  straight 
facet edge, where the supersaturation is fixed
to a value $\sigma_+$. We also assume that $\sigma$
decreases monotonously up to a value $\sigma_-$
at some distance $d$ from the edge.
We therefore assume that the supersaturation that depends on the partial coordinate $y$ as
\begin{eqnarray}
\sigma=\sigma_+ \varsigma(y/d),
\end{eqnarray}
with $\varsigma(0)=1$ and $\varsigma'(y)<0$. 
We also impose that $\varsigma(\vartheta)=\sigma_-/\sigma_+<1$, where $\vartheta$ is a fixed number
so that the supersaturation is equal to $\sigma_-$ for $y=\vartheta d$.
We would then like to evaluate the nucleation rate 
per unit length of facet edge in the direction $x$
\begin{eqnarray}
J_x=\int_0^{\vartheta d} {\mathrm d}y\; J_c(\sigma)\,{\rm e}^{-\sigma_c/\sigma}.
\end{eqnarray}

Since we assume a monotonically decreasing supersaturation profile,
we can change variables and integrate 
over $s=\sigma/\sigma_+$.
In the presence of an essential singularity of the form ${\rm e}^{-b/s}$ when $s\rightarrow 0$,
and for any function $g(s)$ that is finite and regular at $s=1$ (with possible algebraic divergence at $s\rightarrow0$),
we have:
\begin{eqnarray}
\label{eq:large_b}
\int_{\sigma_-/\sigma_+}^{1} {\mathrm d}s\; g(s)\,{\rm e}^{-b/s}\approx g(1)\,\frac{\mathrm e^{-b}}{b}.
\end{eqnarray}
when $b\gg 1$ and $b\gg 1/(1-{\sigma_-}/{\sigma_+})$.
Hence, under the conditions 
\begin{eqnarray}
\label{eq:cond_1}
\sigma_+&\ll&\sigma_c\\
\label{eq:cond_2}
\sigma_+-\sigma_-&\gg& \sigma_+^2/\sigma_c
\end{eqnarray}
the nucleation rate will be dominated by the behavior close to $y=0$.
Since the boundary condition $\varsigma(0)=1$ implies $\varsigma^{-1}(1)=0$, we obtain
\begin{eqnarray}
J_x
=\frac{d}{-\varsigma'(0)}\,
\frac{\sigma_+}{\sigma_c}\,J_+
=\frac{1}{-\frac{\mathrm d\sigma}{\mathrm d y}|_{y=0}}\,
\frac{\sigma_+^2}{\sigma_c}\,J_+
\end{eqnarray}
where $J_+=J|_{\sigma=\sigma_+}$.

A similar analysis also allows one to determine the typical distance $y$
from the edge at which nucleation events should be observed
for a given supersaturation profile:
\begin{eqnarray}
\langle y\rangle_{nuc}=\frac{1}{J_x}\int_0^{\vartheta d} {\mathrm d}y\;y\, J_c(\sigma)\,{\rm e}^{-\sigma_c/\sigma}.
\end{eqnarray}
Using again the same strategy as for Eq.(\ref{eq:large_b}), 
but now with a function $g(s)$ that vanishes for $s=1$, i.e. $g(1)=0$, 
we have for $b\gg 1$:
\begin{eqnarray}
\label{eq:large_b_deriv}
\int_{\sigma_-/\sigma_+}^{1} {\mathrm d}s\; g(s)\,{\rm e}^{-b/s}\approx -g'(1)\,\frac{\mathrm e^{-b}}{b^2}.
\end{eqnarray}
leading to
\begin{eqnarray}
\langle y\rangle_{nuc}
&=&  \frac{d}{-\varsigma'(0)}\frac{\sigma_+}{\sigma_c}
=\frac{\sigma_+^2}{-\frac{\mathrm d\sigma}{\mathrm d y}|_{y=0}\sigma_c}
\end{eqnarray}
This leads to the simple formula
\begin{eqnarray}
J_x=\langle y\rangle_{nuc} J_+.
\end{eqnarray}

When nucleation is confined along the 
edge of the facet, the total nucleation rate is equal to $P J_x$ where $P$
is the perimeter of the facet edge.
Since the rate of nucleation is constant as long as nucleation
does not occur, the probability of having no nucleation event
up to the time $t$ is Poissonian:
\begin{eqnarray}
{\cal P}(t)=\mathrm e^{-\int_0^t\mathrm d \tau PJ_x}
\end{eqnarray}
Note that here, the time $t$ starts  at a conventional time $t=0$
when the system is reset to a reference state. 
As a consequence, the probability density $Q(t)$ that the first
nucleation occurs at time $t$ is 
\begin{eqnarray}
Q(t)=-\partial_t{\cal P}(t).
\end{eqnarray}
The average time for nucleation to occur is then evaluated as
\begin{eqnarray}
\langle t\rangle_{nuc}&=&\int_0^\infty \mathrm d t \;t\;Q(t)
\nonumber\\
&=&\int_0^\infty \mathrm d t \;t\;  PJ_x \mathrm e^{-\int_0^t\mathrm d \tau PJ_x}.
\label{eq:nuc_time_SuppMat}
\end{eqnarray}

The expected value of the position
of the nucleation event when it occurs is
\begin{align}
\label{eq:yloc_SupMat}
\langle y \rangle_{loc}&=\int_0^\infty \mathrm d t\int_0^\infty \mathrm d y
\; y \frac{J(y,t)}{J_x(t)} Q(t)
\nonumber \\
&= \int_0^\infty \mathrm d t \frac{J_x(t)^2P}{J_+}  \mathrm{e}^{-\int_0^t d\tau PJ_x(\tau)}.
\end{align}

We now apply Eqs.(\ref{eq:yloc_SupMat},\ref{eq:nuc_time_SuppMat}) 
to specific forms
of the supersaturation profile.

\paragraph{\bf Localization with pre-existing steps
on the facet.}
As a first example, we aim to mimic 
a situation where a cavity
is forming in the center of the facet.
The steps at the edge of the cavity maintain the supersaturation to a lower value $\sigma_-$.
Hence,  $d$ is constant, or varies slowly. 
Assuming a steady-state
saturation profile, we obtain a linear 
decrease of $\sigma$
from $\sigma_+$ at $y=0$ to $\sigma_-$ at $y=d$:
\begin{eqnarray}
\sigma=\sigma_+-(\sigma_+-\sigma_-)\frac{y}{d}
\end{eqnarray}
 We then have
\begin{eqnarray}
\varsigma(u)=1-(1-\frac{\sigma_-}{\sigma_+})u 
\end{eqnarray}
and $\vartheta=1$. As a consequence, $\varsigma'(u)=-(1-\sigma_-/\sigma_+)$,
leading to
\begin{eqnarray}
J_x
&=&\frac{d}{\sigma_+-\sigma_-}\,\frac{\sigma_+^2}{\sigma_c}\,J_+
\\
\langle y\rangle_{nuc}&=& \frac{d}{\sigma_+-\sigma_-}\,\frac{\sigma_+^2}{\sigma_c}
\end{eqnarray}

If $d$ varies slowly in time (i.e. varies at a time-scale larger than the time
between two nucleation events), then from Eq.(\ref{eq:nuc_time_SuppMat}), we have
\begin{eqnarray}
\langle t\rangle_{nuc}
&=&\frac{1}{P J_x}
\end{eqnarray}
We also find average distance of the nucleation event 
\begin{align}
\langle y \rangle_{loc}= \langle y \rangle_{nuc}
\end{align}
As expected, the since the configuration is time-independent,
the location of the first nucleation event $\langle y \rangle_{loc}$ is the 
same as the location  $\langle y \rangle_{nuc}$ of a nucleation event that would occur if
nucleation occurs at any arbitrary time $t$.

These results suggest that nucleation
is still be localized after the formation
of the cavity in the center of the facet.
Assuming a bunch of steps at a distance of about $d=50\mu$m
from the edge creating a zone where $\sigma_-\approx 0$,
and using $\sigma_+\approx 0.05$
and $\sigma_c=1$, we obtain 
$\langle y \rangle_{loc}\approx 2.5\mu$m.
This is consistent with experimental observations.

\paragraph{\bf Diffusion-limited relaxation of the superstaturation in the absence of other steps.}

As a second example, we consider a supersaturation profile that results from
the relaxation of the supersaturation profile by diffusion 
after the passage of a constant velocity step.
We assume that the initial state is  
a homogeneous supersaturation 
$\sigma_-$ for all $y>0$ at $t=0$. 
At $y=0$, the supersaturation is assumed to be constant 
and equal to $\sigma_+>\sigma_-$ at all times.
The solution of the diffusion problem provides:
\begin{eqnarray}
\label{eq:diffusion_limited_relaxing_supersaturation}
\sigma&=& (\sigma_+-\sigma_-)\mathrm{erfc}(y/(4Dt)^{1/2})+\sigma_-.
\end{eqnarray}
This leads to
\begin{eqnarray}
\varsigma(u)&=&(1-\frac{\sigma_-}{\sigma_+}) \mathrm{erfc}(u)+\frac{\sigma_-}{\sigma_+}
\\
d &=& (4Dt)^{1/2}
\end{eqnarray}
and $\vartheta=\infty$, i.e. the supersaturation decreases to zero at $y\rightarrow\infty$.
In order to use our model based on the nucleation rate (\ref{eq:nucl_theory}), we need to assume that a steady-state for the distribution of monolayer island sizes explored by thermal fluctuations is reached with a time-scale that is faster than that of the evolution of the supersaturation.
This assumption is not valid here, and our model therefore only provides a lower bound for the nucleation time. However, we expect our approach to catch the main features of the localization of nucleation.

Using equations (\ref{eq:nuc_time_SuppMat},\ref{eq:yloc_SupMat}), we now  obtain:
\begin{eqnarray}
J_x
&=&(\pi Dt)^{1/2} \frac{\sigma_+^2}{(\sigma_+-\sigma_-)\sigma_c}\,J_+.
\\
\label{eq:ynuc_constant_veloc}
\langle y\rangle_{nuc}&=& 
(\pi Dt)^{1/2} \frac{\sigma_+^2}{(\sigma_+-\sigma_-)\sigma_c}
\end{eqnarray}
Assuming that the variation of the perimenter $P$ and the supersaturation $\sigma_+$
and $\sigma_-$ are slower than the variation of $d$, we find
\begin{eqnarray}
\label{eq:nuc_time_recov_SupMat}
\langle t\rangle_{nuc}
&=&
 \frac{\Gamma[{5/3}]}{ (\pi D)^{1/3}}
\left( \frac{3}{2} \frac{(\sigma_+-\sigma_-)\sigma_c}{\sigma_+^2 J_+P}\right)^{2/3}
\nonumber \\
\end{eqnarray}
We have fitted  to the experimental data in Figure 2 using the expressions (\ref{eq:nucl_theory}) and (\ref{eq:ynuc_constant_veloc})
with $\sigma_c$ and $\rho_s$ as free parameters.  
We obtain  $\Omega_2\rho_s=2.8\times 10^{-6}$.
In addition, we find $\sigma_c\approx 1.4$,
which leads to $\Gamma\approx 0.44$~nm.
These values are close to those obtained
using the expression for a homogeneous nucleation
rate on the facet.

We also obtain the position $\langle y \rangle_{loc}$
of the nucleation event as:
\begin{align}
\langle y \rangle_{loc}
&=
\left(\frac{\pi D}{J_+P}\right)^{1/3}
\left(\frac{\sigma_+^2}{(\sigma_+-\sigma_-)\sigma_c}\right)^{2/3}
\frac{3^{1/3}}{2^{1/3}}\Gamma[\frac{4}{3}]
\end{align}
Hence
\begin{align}
\frac{\langle y \rangle_{loc}}{\langle y \rangle_{nuc}|_{t=t_{nuc}}}
=\frac{\Gamma[\frac{4}{3}]}{\Gamma[\frac{5}{3}]^{1/2}}=0.9398..
\end{align}
Since this ratio is close to $1$, we conclude that 
the detailed description of the time-dependence of 
the position of the nucleation event via the time-dependence 
of $d$ does not bring a significiant quantitative correction
to the average position of nucleation.

We can also rewrite the result as
\begin{align}
\label{eq:yloc_recovery_SM}
\langle y \rangle_{loc}
&= (\pi D \langle t \rangle_{nuc})^{1/2}
\frac{\sigma_+^2}{(\sigma_+-\sigma_-)\sigma_c}
\frac{\Gamma[\frac{4}{3}]}{\Gamma[\frac{5}{3}]^{1/2}}
\end{align}

Quantitatively, we find that $d$ varies from $50\mu$m
in $1$s to about $150\mu$m in $10$s.
Assume that the supersaturation has been depleted
by the passage of a 
step  in a film of thickness $30$nm. 
Then from Eq.(\ref{eq:sigma<_appendix}),
we obtain $\sigma_+-\sigma_-=1/\Theta_{eq}\approx0.036$. 
Note that this difference
is smaller, but of the same order as the supersaturation
imposed at the boundary in the single-nucleation regime 
$\sigma_+\approx 0.05$.
We finally obtain from Eq.(\ref{eq:yloc_recovery_SM})
$\langle y \rangle_{loc}=3\mu$m
for a nucleation time $\langle t\rangle_{nuc}\approx 1$s,
and 
$\langle y \rangle_{loc}=9\mu$m
for a nucleation time $\langle t\rangle_{nuc}\approx 10$s.

\subsection*{Step front instability}
\begin{figure*}
    \centering
    \includegraphics[width=12cm]{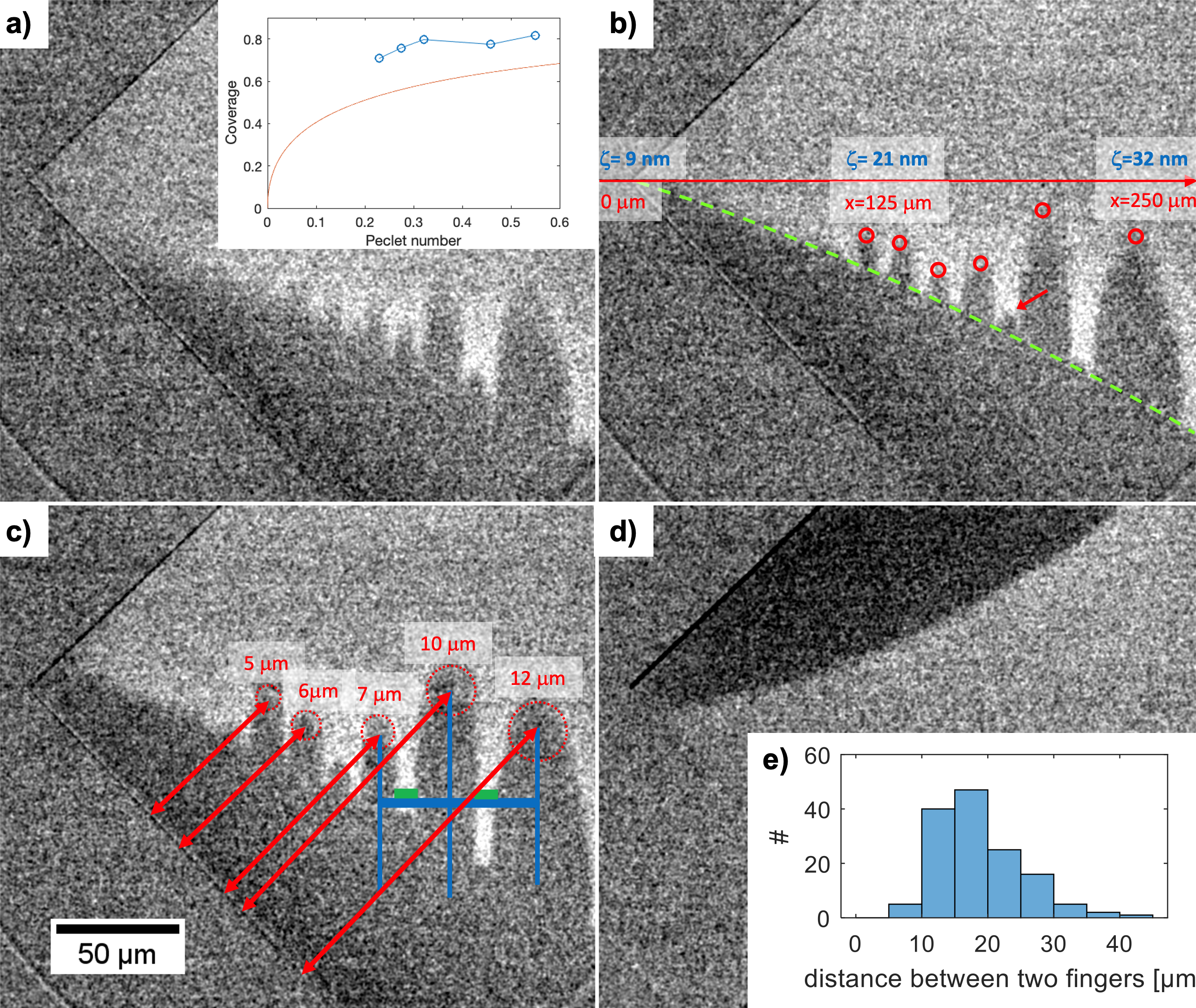}
   \caption{Instability of step flow. {\bf a)-c)} consecutive average subtracted RICM images at 0.45 s interval of step flow at an edge of a confined NaClO$_3$ interface showing instability of the front. The dark areas correspond to a smaller distance to the confining glass and thus to the newly formed layer.
Tracking of tips and step front base is indicated in b). The crystal is tilted along the red line in b), the glass-crystal distances $\zeta$ are given in blue for the distances (in red) along the red arrow. In c) we indicate the measurements of finger tip radii, where the distance to the crystal edge is measured (red arrows) and the measurement of distance between finger midpoints (blue lines) and width of zone that is not covered by the fast, instable front (green).
d) Slow double layer front at the same interface and under the same conditions directly after a)-c) not showing instabilities. e) Histogram of distance of adjacent fingers with a mean value of 18.6~nm and a standard deviation of 6.4~nm. Frame rate: $1/dt=1/450$~ms, integration time: $t_{ac}=450$~ms, supersaturation in the bulk:
$\sigma=0.053$.
Inset in a) is a comparison between the theoretical relation between Peclet number and coverage from equation(\ref{eq:Ivantsov_SupplMat}) and the corresponding parameters from measurements in these images. }
    \label{fig:S5}
\end{figure*}

\subsubsection*{Experiments}
Several experiments showed unstable step fronts. In most cases the crystals were too small to show well developed fingers that lend themselves to quantitative analysis. We therefore concentrate the discussion on the experiment shown in movie S4 and Figs.~4 and S5 where the bulk supersaturation is 0.053 and the $\bar{\zeta}=20$~nm.
The first four consecutive step fronts pass at time intervals of 54$\pm$7~s and develop  finger-like instabilities. All 4 unstable fronts travel in a direction between the two maximum kinetic anisotropy directions (see arrows of kinetic anisotropy in Fig.~4~A-D). The fifth step front (Fig.~S5~D) arrives only 9 seconds after the preceeding front and travels in the low velocity direction. The time between fronts and the direction of travel are therefore important to whether the fronts destabilize or not.

The time $t$ between two fronts determines how far diffusion has transported ions along the confined crystal surface. The diffusion length $l=\sqrt(Dt)$ is 180~$\mu$m for the unstable fronts and 70~$\mu$m for the last front. No part of the images are more than 160~$\mu$m from a crystal edge. The fact that the last front starts out with a velocity of 18~$\mu$m/s and slows down with the square root of time is consistent with the front moving in a diffusion controlled concentration field.

Fig.~4A-C follow the progression of one of the unstable fronts. In Fig.~S5~B the finger tips are indicated with red circles and the slow part of the front is indicated with a dashed green line. The finger tips propagate at constant speed and the slow part propagates at a velocity slowing down with the square root of time. This indicates that the fingers propagate by using the ions already present in the confined fluid film whereas the slow part depends on diffusion transport of ions from the bulk to propagate. The fraction of the area covered by the fingers before the slow front arrives is 0.7-0.9. One interpretation of this finite coverage by the fast growing layer is that it is limited by the number of ions already present in the fluid film. The ion coverage in the confined fluid is $\Theta_{eq}\sigma=\zeta c_0\sigma/(z_0c_s)=\zeta\sigma/1.2$~nm, where $c_0/c_s=0.72/2.54$ and $z_0=0.33$~nm. 

\subsubsection*{Theoretical considerations}
Let us consider a straight and isolated atomic step along the $x$ axis 
moving at velocity $V$ in the $y$ direction during growth.
If small perturbations $\xi(x,t)$ of the step position along
$y$ are amplified, then the step is unstable.
When the perturbation is small enough, the dynamical
equations governing the motion of the step can be analyzed in perturbations,
and to leading order one obtains a set of linear equations governing 
the evolution of $\xi(x,t)$.
Assuming for simplicity that the step velocity is constant in time,
a single Fourier mode of the perturbation $\xi_q(t)$ of wavevector $q$ 
therefore grows or decays exponentially in time as ${\rm e}^{i\omega t}$. The linear equations for $\xi_q(t)$
then provide a relation between $i\omega$ and $q$,
which is called the dispersion relation.
If $\Re e[i\omega]>0$ for some values of $q$, then the 
perturbations $\xi_q(t)$ will grow exponentially in time, and
the step will be unstable. 

In the long wavelength limit,
the dispersion relation contains two terms. The first
term is destabilizing and is proportional to the 
velocity $V$, and the second term is a stabilization
term due to line tension effects
\begin{eqnarray}
\label{eq:dispersion_MS}
i\omega=Vq-2\tilde\Gamma \Theta_{eq} D q^3
\end{eqnarray}
where $D$ is a diffusion constant,
and $\tilde\Gamma=\Omega\tilde\gamma/nk_BT$ is the so-called capillary length,
where $\Omega$ is the molecular area in a solid layer,
$n=2$ is the number of ions in the liquid for one solid molecule, 
 $k_BT$ is the thermal energy
and $\tilde\gamma$ is the step stiffness.
The stiffness is in general a function of the orientation
angle $\phi$, and is related to the step free energy $\gamma(\theta)$
via the relation $\tilde\gamma(\theta)=\gamma(\theta)+\gamma''(\theta)$.

From the condition $i\omega>0$, we find
that the perturbations with wavelength larger than 
\begin{eqnarray}
\label{eq:lambda_0_paper}
\lambda_0=2\pi(2\tilde\Gamma \Theta_{eq}\ell)^{1/2},
\end{eqnarray}
where $\ell=D/V$ is usually called the diffusion length,
are unstable. The wavelength
of the most unstable mode, i.e., the mode with the largest $\Re e[i\omega]$, reads
\begin{eqnarray}
\lambda_m=3^{1/2}\lambda_0=2\pi(6\tilde\Gamma \Theta_{eq} \ell)^{1/2}.
\end{eqnarray}
In the isotropic case, we have $\tilde\gamma=\gamma$ and $\tilde\Gamma=\Gamma/n$,
leading to the expression of the main text.

When the step is unstable, small perturbations grow until
nonlinearities come into play. These nonlinearities control
the emerging morphology. A number of studies have focused
on similar problems in two or three dimensions.

The driving force is measured by the coverage $\Theta_{eq}\sigma=\zeta c_0\sigma/(z_0c_s)=\zeta\sigma/1.2$~nm, where $c_0/c_s=0.72/2.54$, $c_s$ is the molar density of the solid, $\zeta$ is the liquid film thickness and $z_0$ is the height of a monolayer. This number measures the number of monolayers of solid that
can be formed from the excess of ions in the supersaturated liquid. Considering a supersaturation $\sigma\approx 0.05$,
we obtain $\Theta_{eq}\sigma\approx \zeta/22$~1/nm where 
the film thickness $\zeta$ is in nanometers.

If $\Theta_{eq}\sigma>1$, i.e.
in our experiments if $\zeta >22$nm, 
there is enough material in the liquid
for a straight step to grow at constant velocity into a supersaturation $\sigma$.
The supersaturation decreases from the value $\sigma$ far in front of the step
to the value $\sigma_<$
at the step and behind it, 
which is depleted by an amount that corresponds to
the formation of the new monolayer:
\begin{eqnarray}
\label{eq:sigma<_appendix}
\sigma_<=\sigma-\frac{1}{\Theta_{eq}}
\end{eqnarray}
The step velocity then reads
\begin{eqnarray}
V = \alpha k(\theta)\sigma_< 
\end{eqnarray}
where the kinetic coefficient $\alpha k(\theta)$ is defined in the main text.

Linear stability analysis suggest that 
these steps undergo a Mullins-Sekerka instability when
\begin{eqnarray}
\label{e:summary_straight_step_ss}
\Theta_{eq}\sigma  < 1 + \frac{D}{\alpha k(\theta) \tilde\Gamma}
\end{eqnarray}
The nonlinear dynamics has only been investigated
theoretically close to the threshold \cite{Misbah2010}, suggesting that isotropic
steps undergo spatio-temporal chaos governed by the Kuramoto-Sivashinsky equation. 
However, in our experiments $\Theta_{eq}\sigma$
is at most $\sim 5$
while $nD/(\bar\alpha \Gamma)\sim 10^3$.
This correspond to a regime which is far from the threshold,  where $q_0\ell\gg 1$,
with $q_0=2\pi/\lambda_0$. In this limit, the usual
expression (\ref{eq:lambda_0_paper}) of $\lambda_0$ is valid,
but to our knowledge, nothing is known about nonlinear dynamics.

In the other regime when the coverage is low $\Theta_{eq}\sigma<1$, corresponding to small film thicknesses $\zeta<1.2/\sigma$~nm,
there is not enough ions in the liquid film to form a monolayer via the motion of a step.
As a consequence, ions have to diffuse from an increasing distance to
be incorporated in the step as time goes forward. This results 
in a decreasing step velocity 
$v_{step}\sim (D/t)^{1/2}$.

This regime gives rise to the celebrated "dendrite" morphologies.
Dendrites correspond to a constant-velocity parabolic step profile,
which possibly undergoes side-branching away from the tip.
The growth direction and the tip radius are controlled by anisotropy.
In contrast, when anisotropy is negligible, one finds seaweed shapes characterized by permanent branching due to the splitting of their tips.

Irrespective of anisotropy, the tip of the parabolic Ivanstov dendrite solution~\cite{Kassner1996}
is constrained by a relation between  the tip radius $R$ and the tip velocity $v_{tip}$:
\begin{eqnarray}
\label{eq:p_def_SupplMat}
p=\frac{Rv_{tip}}{2D}.
\end{eqnarray}
where $p$ is a dimensionless number called the Peclet number,
which exhibits a non-trivial dependence on the coverage
\begin{eqnarray}
\label{eq:Ivantsov_SupplMat}
\Theta_{eq}\sigma = (\pi p)^{1/2} \mathrm e^p \mathrm{Erfc}[p^{1/2}]
\end{eqnarray}
Inset in Fig.~S5~a) is a comparison between this relation between Peclet number and coverage and the corresponding parameters from measurements. In Fig.~S5 we indicate the measurements or $R$ for each of the fingers. Using $v_{tip}=55\mu$m and $D=0.6\cdot 10^{-9}$m$^2$/s we obtain the experimental Peclet numbers $p$ from equation (\ref{eq:p_def_SupplMat}).

The time scale of motion of the fingers is short compared to the long time scale of the diffusion controlled evolution of supersaturation in the thin film. We may estimate the coverage $\Theta_{eq}\sigma$ at the time of passage of the fingers in two different ways: by solving the diffusion equation or by measuring the fraction of the area covered by the fingers.

Because the crystal is tilted and 
because of the finite time since the last layer grew the coverage at each finger tip is a function of its position. The supersaturation can be estimated as a function of distance to the nearest edge (as indicated in Fig~S5) and the time $t$ since the last growth front passed:
  \begin{equation}
\sigma(x,t)/\sigma(x=0)=\mathrm{erfc}\left(\frac{x}{2\sqrt(Dt)}\right).   
  \end{equation}
  This assumes that $\sigma(t=0)=0$ and is not corrected for the tilt of the crystal. In order to calculate the coverage $\Theta_{eq}\sigma=\zeta\sigma/(1.2$~nm$)$ the glass-crystal distance $\zeta$ is evaluated for each tip. The resulting coverages in the range 0.7-0.8 are plotted in Fig.S5~a.  For one of the fingers we may confidently estimate the coverage by the fraction of the crystal area the fingers cover before the slow front arrives. Fig.~S5~c shows this measurement of coverage $\Theta_{eq}\sigma\approx 0.7$ for the second rightmost finger. This is 10\% smaller than estimated from the diffusion calculation. In the inset of Fig.~S5~a) we plot the coverage (from the diffusion calculation) for each finger tip as function of Peclet number from (\ref{eq:p_def_SupplMat}). This agrees very well with the prediction from equation (\ref{eq:Ivantsov_SupplMat}).

Two-dimensional numerical simulations of the dynamics with coverage lower than one
has been reported in many papers (see for example~\cite{Ben-Jacob1993}).
The results show dendrites and seaweed features at small values
of the coverage. However, when the coverage approaches 1,
a continuous front emerges with a porous monolayer.
As a consequence, fingering should actually only be observed
approximatively for $\Theta_{eq}\sigma< 0.9$. The porosity
of the newly formed monolayer decreases as one approaches unit coverage.

\section*{Calcite experiments}
\begin{figure}
    \centering
    \includegraphics[width=8cm]{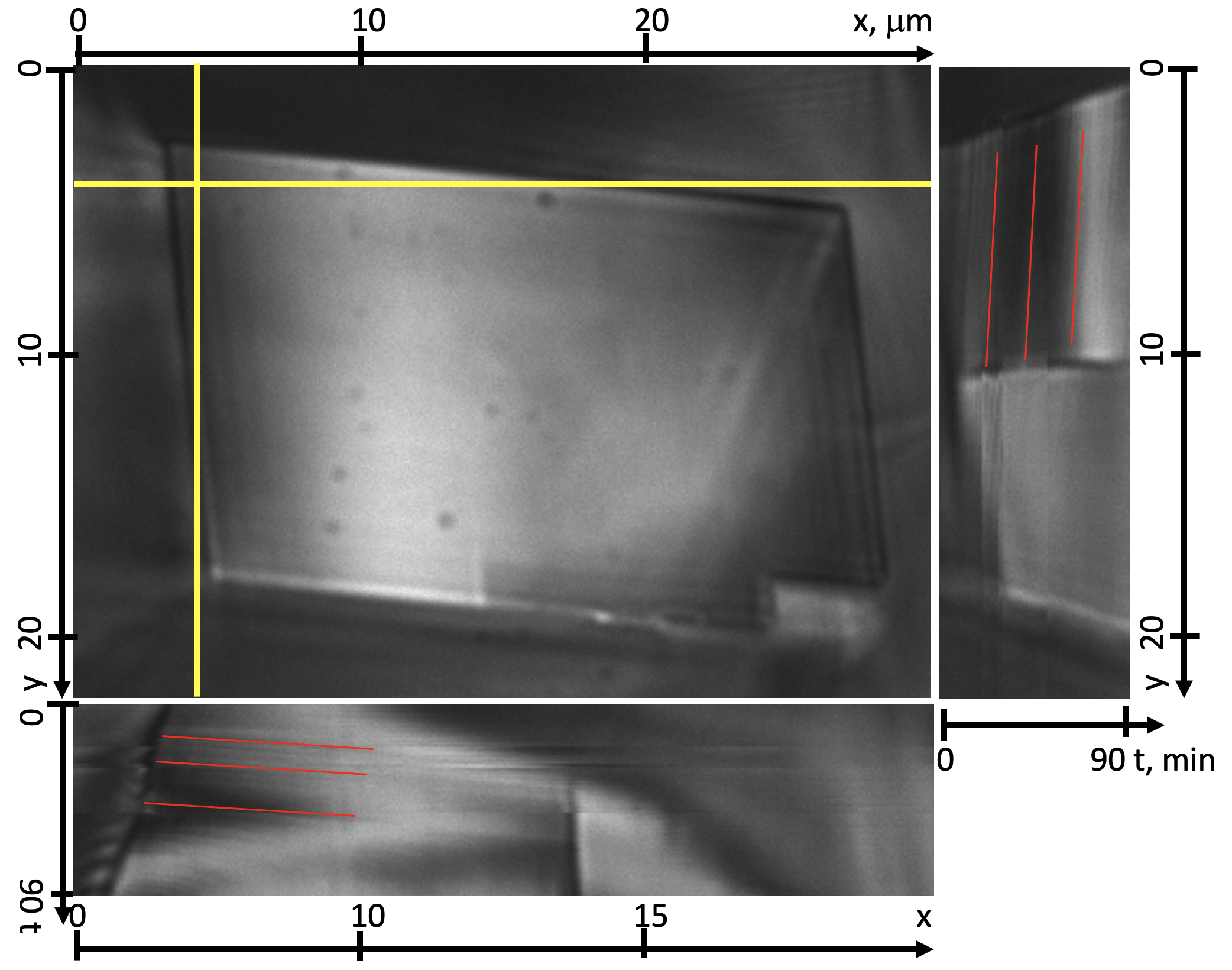}
    \caption{Surface of calcite crystal 30~nm above confining glass surface in water with 0.8~mM CaCO$_3$ concentration ($\sigma=0.6$) imaged once every minute. Bottom and right plot show intensities along yellow line as function of time. The red lines trace the motion of waves of step flow during growth of the confined calcite surface. The slope of the red lines yield the step flow velocity in the x- and y-direction.}
    \label{fig:calcite}
\end{figure}
The calcite experimental setup has been described in detail elsewhere~\cite{Li2021}. A calcite crystal (see Figure~\ref{fig:calcite}) of approximately 15x25~$\mu$m is kept at a CaCO$_3$ concentration of 0.801$\pm$0.002~mM, which corresponds to a supersaturation of $\sigma=0.6$ and a saturation index of $\Omega$=0.44~\cite{Li2017b}.  

Supplementary movie SM6 shows "waves" of high and low intensity move along the rim from bottom left, around the top left corner and on to the top right. These intensity variations are due to changes in the glass-crystal distance $\zeta$ as molecular steps move across the surface. The step flow emanates from a step edge on the rim in much the same way as from the dislocation source in Figure~2.

From \cite{Li2021} we know that when the distance between the glass surface and the calcite surface was 30~nm the vertical growth rate was 2.6~nm/min. The step height on calcite is 0.34~nm and the vertical growth rate of 8 layers per minute corresponds to a horizontal single molecular step spacing of 170-340~nm. This is well below the horizontal resolution limit of the objective in this experiment and the waves of step flow observed in movie SM6 correspond to several molecular steps, either equidistant or joined in step bunches. The growth regime on this calcite surface is not a "single step" regime as for several of the NaClO$_3$ experiments. Thus, even though we cannot resolve single steps, the interference contrast allows us to measure the "collective" speed of the molecular steps along the surface. 

Figure~\ref{fig:calcite} shows the same crystal at time $t=0$ with a vertical and a horizontal yellow line that indicate where intensity has been measured and displayed as function of time in the two side images. The motion of the intensity change caused by the flow of molecular steps manifests as dark and bright lines in the $x-t$ and $y-t$ images in Figure~\ref{fig:calcite} and the slope of the lines measure the velocity of the molecular steps. We have highlighted some of these lines with red in Figure~\ref{fig:calcite}. The step flow velocities are determined to be 43$\pm$10~nm/s in the x-direction and 22$\pm$5~nm/s in the y-direction. This is one order of magnitude faster than determined from AFM measurements\cite{Teng2000,Bracco2013} at similar reported saturation index. We have, however shown that calcite growth rates as function of saturation index measured by AFM are 2 orders of magnitude too small, probably due to insufficient control of the saturation index at the calcite surface~\cite{Li2018_CGD}. Since the supersaturation at the confined surface decreases with distance from the crystal edge we find our present measurement $v_s/\sigma=40-70$~nm/s in agreement with previous findings.

\bibliography{stepFlowBib}